\documentclass[aps,pra,twocolumn,a4paper,reprint,noeprint]{revtex4-1}
\usepackage[utf8]{inputenc}
\usepackage{graphicx}
\usepackage{amsmath}
\usepackage{amsfonts}
\usepackage{amssymb}
\usepackage[usenames,dvipsnames,svgnames,table]{xcolor}
\usepackage[USenglish]{babel}
\usepackage{hyperref}
\usepackage{grffile}
\usepackage{color}

\newcommand{\ignore}[1]{}

\allowdisplaybreaks

\begin{document}

\title{Experimental consequences of Bogoliubov Fermi surfaces}

\author{Clara J. Lapp}
\affiliation{Institute of Theoretical Physics, Technische Universit\"at Dresden, 01069 Dresden, Germany}

\author{Georg B\"orner}
\affiliation{Institute of Theoretical Physics, Technische Universit\"at Dresden, 01069 Dresden, Germany}

\author{Carsten Timm}
\email{carsten.timm@tu-dresden.de}
\affiliation{Institute of Theoretical Physics, Technische Universit\"at Dresden, 01069 Dresden, Germany}

\date{September 27, 2019}

\begin{abstract}
Superconductors involving electrons with internal degrees of freedom beyond spin can have internally anisotropic pairing states that are impossible in single-band superconductors. As a case in point, in even-parity multiband superconductors that break time-reversal symmetry, nodes of the superconducting gap are generically inflated into two-dimensional Bogoliubov Fermi surfaces. The detection and characterization of these quasiparticle Fermi surfaces requires the understanding of their experimental consequences. In this paper, we derive the low-energy density of states for a broad range of possible nodal structures. Based on this, we calculate the low-temperature form of observables that are commonly employed for the characterization of nodal superconductors, i.e., the single-particle tunneling rate, the electronic specific heat and Sommerfeld coefficient, the thermal conductivity, the magnetic penetration depth, and the NMR spin-lattice relaxation rate, in the clean limit. We also address the question whether the topological invariant of the Bogoliubov Fermi surfaces is associated with topologically protected surface states, with negative results. This work is meant to serve as a guide for experimental searches for Bogoliubov Fermi surfaces in time-reversal-symmetry-breaking superconductors.
\end{abstract}

\maketitle

\section{Introduction}
\label{sec.intro}

The common view of superconductors evokes a condensate of Cooper pairs formed by electrons with a single internal degree of freedom---the spin. The formation of this condensate is understood as an instability of the normal-state Fermi surface, which may consist of a single or multiple sheets. In the latter case, the superconducting pairing amplitude is typically different on different sheets but beyond this, the presence of multiple bands does not have any qualitative consequences. In this picture, the pairing amplitude can have zeros in momentum space, called \emph{nodes}. These nodes are either points or lines in the three-dimensional Brillouin zone. Our interest here is in centrosymmetric, even-parity superconductors. For these materials, the pairing amplitude generically has line nodes in the presence of time-reversal symmetry (TRS) and point nodes if TRS is spontaneously broken \cite{SiU91}. TRS-breaking states can also have line nodes in high-symmetry planes.

It has recently been realized that this picture is incomplete. The pairing of electrons with additional internal degrees of freedom resulting, e.g., from different orbitals or basis sites, can lead to qualitatively new pairing states. Such pairing states have for example been proposed for iron-based superconductors \cite{GSZ10,NGR12,NKT16,OCS16,CVF16,NYS17,ASO17,SBK19}, $\mathrm{Cu}_x\mathrm{Bi_2Se_3}$ \cite{Fu14,YTN17}, cubic systems such as half-Heusler compounds \cite{BWW16,TSA17,YXW17,SRV17,YuL18,BoH18,KWN18,KOK18,RGF19,SMP19}, $\mathrm{UPt}_3$ \cite{NoI16,Yan16}, transition-metal dichalcogenides \cite{OYK18,MoK18}, and twisted bilayer graphene \cite{GZF18,SuL18,Wu19}. It has been shown that in centrosymmetric multiband superconductors that break TRS, point and line nodes are generically replaced by spheroidal and tubular Fermi surfaces of Bogoliubov quasiparticles, due to interband pairing \cite{ABT17,BAM18}. We speak of ``inflated'' point and line nodes, respectively, in the following. These two-dimensional \emph{Bogoliubov Fermi surfaces} (BFSs) are protected by a topological $\mathbb{Z}_2$ invariant \cite{KST14,ZSW16}, which can be expressed in terms of a Pfaffian~\cite{ABT17,BAM18}.

\begin{table*}[tb]
\begin{ruledtabular}
\begin{tabular}{cccccc}
  & \multicolumn{2}{c}{\hspace{-2.7em}$g=2$ (point nodes)} & \multicolumn{2}{c}{\hspace{-3.3em}$g=1$ (line/double-Weyl nodes)} & \\
\raisebox{1.5ex}[-1.5ex]{observable} & uninflated & inflated & uninflated & inflated &
  \raisebox{1.5ex}[-1.5ex]{full gap} \\ \hline
$\Delta\lambda$ & $T^2$ & $T^2$ & $T$ & $T\, e^{-h/k_BT}$ &
  $\displaystyle \frac{\Delta^{1/2}}{T^{1/2}}\, e^{-\Delta/k_BT}$ \\[1.5ex]
$c$, $\kappa$ & $T^3$ & $\displaystyle h^2\, k_BT + \frac{7\pi^2}{5}\, (k_BT)^3$ &
  $T^2$ & $\displaystyle h\, k_BT + \frac{6}{\pi^2}\, h^2\, e^{-h/k_BT}$ &
  $\displaystyle \frac{\Delta^{5/2}}{T^{3/2}}\, e^{-\Delta/k_BT}$ \\[1.5ex]
$\gamma$ & $T^2$ & $\displaystyle h^2 + \frac{7\pi^2}{5}\, (k_BT)^2$ &
  $T$ & $\displaystyle h + \frac{6}{\pi^2}\, \frac{h^2}{k_BT}\, e^{-h/k_BT}$ &
  $\displaystyle \frac{\Delta^{5/2}}{T^{5/2}}\, e^{-\Delta/k_BT}$ \\[1.5ex]
$\displaystyle \frac{1}{T_1T}$ & $T^4$ & $\displaystyle h^4 + \frac{2\pi^2}{3}\, h^2\, (k_BT)^2$ &
  $T^2$ & $\displaystyle h^2 + 4h\, k_BT\, e^{-h/k_BT}$ &
  $\displaystyle \frac{\Delta}{T}\, \ln\frac{k_BT}{\omega_0}\, e^{-\Delta/k_BT}$
\end{tabular}
\end{ruledtabular}
\caption{Leading terms of the temperature dependence of the temperature-dependent part of the penetration depth, $\Delta\lambda$, the electronic contribution to the specific heat $c$ and the thermal conductivity $\kappa$, the electronic Sommerfeld coefficient $\gamma=c/T$, and the NMR spin-lattice relaxation rate $1/T_1T$ for uninflated and inflated nodes with exponents $g=2$ (linear point nodes) and $g=1$ (linear line nodes and double-Weyl point nodes). For the thermal-conductivity tensor, only the temperature-dependent scalar prefactor is shown. The $g=1$ results in this case only apply to linear line nodes, not to double-Weyl point nodes. The results on $c$, $\kappa$, and $\gamma$ for inflated nodes disregard corrections due to the temperature dependence of the gap. Overall proportionality factors independent of temperature $T$ and pseudomagnetic field $h$ are omitted. For comparison, expressions for a superconductor with constant gap $\Delta$ are given in the last column \cite{PrG06, BCS57, HeS59, SiU91, Tin96, note.full}. For $1/T_1T$, the nonzero nuclear resonance frequency $\omega_0$ must be reintroduced for a constant gap. Since it typically corresponds to temperatures on a order of a few millikelvin, we have assumed $\omega_0 \ll k_BT \ll \Delta$ \cite{note.full}.}
\label{tab.sum}
\end{table*}

In order to be able to detect BFSs, it is paramount to determine their experimental consequences. While angular resolved photoemission spectroscopy (ARPES) would be ideal for mapping out the quasiparticle dispersion, it is often not feasible due to low critical temperature or bad surface quality. Hence, we here consider the specific heat, the thermal conductivity, the magnetic penetration depth, and the NMR spin-lattice relaxation rate. These probes are routinely used to determine the nodal structure of unconventional superconductors since they show characteristic power-law or exponential temperature dependences at low temperatures. We obtain the corresponding low-temperature expansions for inflated point and line nodes with rather general low-energy dispersions, including linear and quadratic point nodes and line nodes with linear dispersion. In this work, we focus on the clean limit, where the energy scale characteristic for the smearing of the density of states (DOS) by disorder is small compared to all other energy scales. For ease of reference, we present the main results in Table~\ref{tab.sum}.

Beyond the bulk properties discussed so far, it is natural to consider surface-bound electronic states since topological invariants are often associated with surface states through a bulk-boundary correspondence. Such surface states could then be probed by tunneling experiments. One example are flat surface bands in noncentrosymmetric superconductors, which should lead to a zero-bias peak in tunneling \cite{TMY10,ScR11,BST11,SBT12,ScB15,TSA17}. However, we are not aware of a mathematical argument for (or against) surface states associated with the $\mathbb{Z}_2$ invariant of BFSs. In this work, we use numerical exact diagonalization of a Bogoliubov--de Gennes Hamiltonian for a slab to search for surface states, with negative results.

The remainder of this paper is organized as follows: In Sec.\ \ref{sec.DOS}, the low-energy form of the quasiparticle DOS is derived for various types of nodes. We also comment on single-particle tunneling, which directly probes the DOS. Based on the results, we obtain low-temperature expansions for the magnetic penetration depth, the electronic contribution to the specific heat, the NMR spin-lattice relaxation rate, and the electronic contribution to the thermal conductivity in Sec.\ \ref{sec.obs}. The question of surface states associated with the topological $\mathbb{Z}_2$ invariant is considered in Sec.\ \ref{sec.surface}. We summarize our results and draw conclusions in Sec.~\ref{sec.summary}.

\section{Density of states}
\label{sec.DOS}

In this section we derive the DOS close to various types of nodes. Our strategy is to expand the quasiparticle dispersion about the Fermi energy (zero by convention), to leading order. This allows us to obtain analytical results that exhibit parametric dependences and exponents of power laws. We start with uninflated and inflated point nodes with a rather general triaxial power-law dispersion and then consider uninflated and inflated line nodes with a power-law dispersion. We shall see that line nodes can be viewed as limiting cases of point nodes.

\subsection{General point nodes}

Neglecting interband pairing to start with, point nodes generically appear in high-symmetry directions for broken TRS. (We do not consider accidental nodes here, which can occur anywhere in the Brillouin zone.) At any such point node, the direction orthogonal to the normal-state Fermi surface is special---in this direction, the superconducting contribution to the quasiparticle dispersion vanishes. The dispersion is thus generically linear to leading order and given by $E_\mathbf{k} = v_F k_\perp$, where $v_F$ is the normal-state Fermi velocity and $k_\perp$ is the momentum component normal to the Fermi surface. The two directions orthogonal to $k_\perp$ are tangential to the normal-state Fermi surface. The quasiparticle dispersion in these directions describes how the superconducting gap opens.

Denoting the momentum relative to an uninflated point node by $\mathbf{q}$ in such a way that $q_3$ describes the direction orthogonal to the normal-state Fermi surface, we expand the quasiparticle dispersion around the node as
\begin{equation}
E^0_\mathbf{q} = \pm \sqrt{\alpha_1^2 |\Delta_0|^2\, |q_1|^{2m}
  + \alpha_2^2 |\Delta_0|^2\, |q_2|^{2n}
  + v_F^2 q_3^2} ,
\end{equation}
where $\alpha_1, \alpha_2 > 0$ are constants, $\Delta_0$ is the global scale of the pairing amplitude, and $2m, 2n > 0$ are exponents that need not be integers. We set $\hbar$ to unity throughout the paper. 

Except at the point nodes, the quasiparticle bands are twofold degenerate, which can be described by a pseudospin $1/2$. Interband pairing generates a pseudomagnetic field $\mathbf{h}$ \cite{ABT17,BAM18}, which couples to the pseudospin and splits the degeneracy of the bands. If interband pairing is small compared to the energy difference between the bands, the former can be treated perturbatively in an effective single-band model close to each Fermi sheet. As shown in Refs.\ \cite{ABT17,BAM18}, the leading term in $\mathbf{h}$ is of second order in the interband pairing and is thus typically small. Furthermore, it is generically nonzero at the nodes and hence can be represented, to leading order, by a constant. The dispersion in the vicinity of the (former) point nodes then becomes
\begin{equation}
E_\mathbf{q} = \pm h \pm \sqrt{\alpha_1^2 |\Delta_0|^2\, |q_1|^{2m}
  + \alpha_2^2 |\Delta_0|^2\, |q_2|^{2n}
  + v_F^2 q_3^2} ,
\label{eq.Eq.3}
\end{equation}
where $h = |\mathbf{h}|$. Here and in the following, the two signs can be chosen independently, giving four bands.

\begin{figure}[tb]
\centerline{\includegraphics[width=0.9\columnwidth]{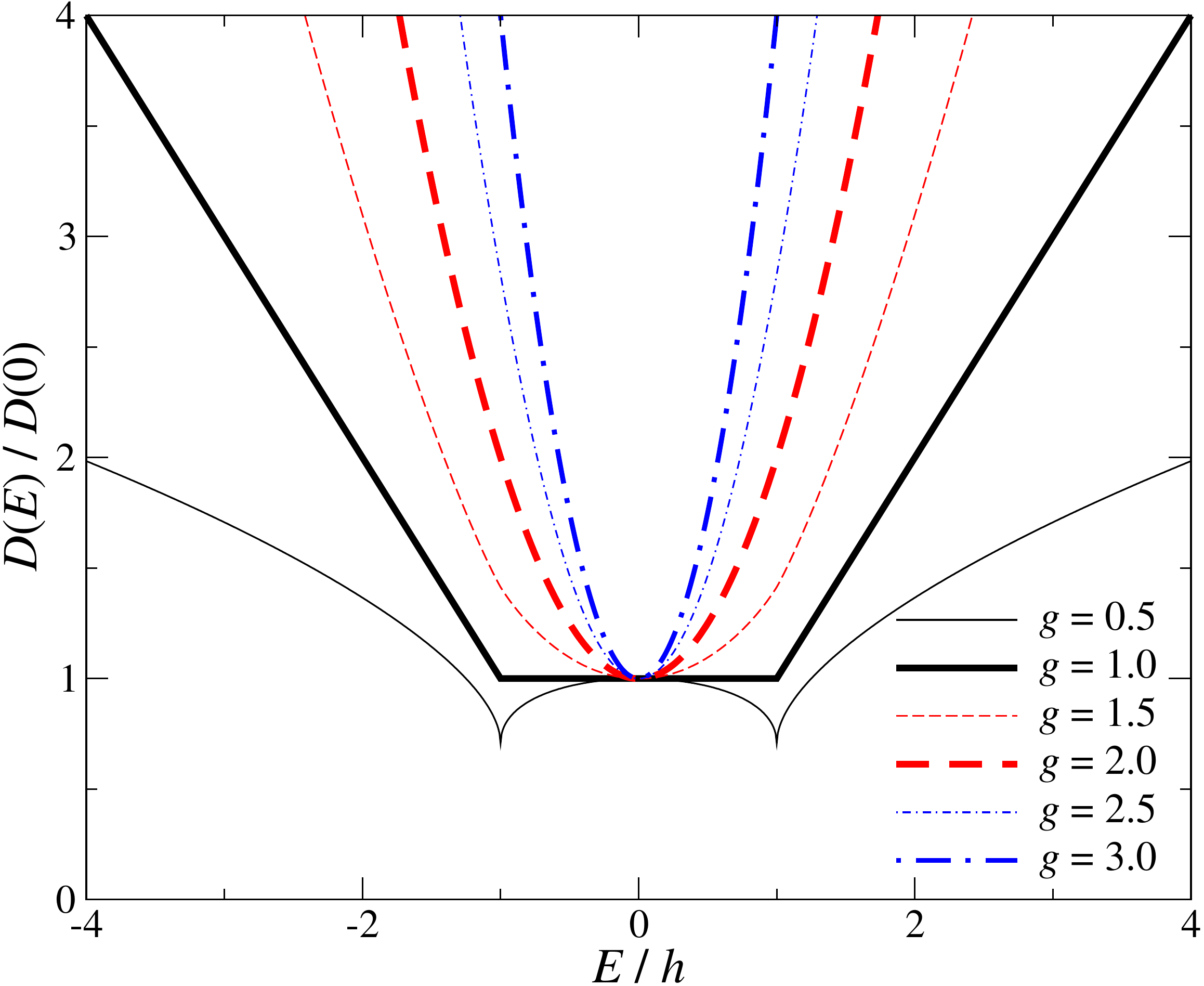}}
\caption{Density of states for various values of the exponent $g=1/m+1/n$.}
\label{fig.DOS}
\end{figure}

The evaluation of the DOS
\begin{equation}
D(E) = \int_{\mathbb{R}^3} \frac{d^3q}{(2\pi)^3}\, \delta(E-E_\mathbf{q})
\label{eq.DOS.point.2}
\end{equation}
is outlined in Appendix \ref{app.DOS.point}, with the result
\begin{align}
D(E) &= \frac{2\sqrt{\pi}}{(2\pi)^3}\,
  \frac{1}{mn\, (\alpha_1|\Delta_0|)^{1/m}\, (\alpha_2|\Delta_0|)^{1/n}\, v_F} \nonumber \\
&\quad {}\times \frac{\Gamma\left(\frac{1}{2m}\right) \Gamma\left(\frac{1}{2n}\right)}
    {\Gamma\left(\frac{1}{2} + \frac{1}{2m} + \frac{1}{2n}\right)} \nonumber \\
&\quad {}\times \big( |E + h|^{1/m+1/n} + |E - h|^{1/m+1/n} \big) \nonumber \\
&\equiv \frac{c_{m,n}}{2}\, \big( |E + h|^g + |E - h|^g \big) ,
\label{eq.DE.g.3}
\end{align}
where
\begin{equation}
g = \frac{1}{m} + \frac{1}{n} > 0
\end{equation}
is a characteristic exponent, which will play an important role. The DOS is the sum of two contributions from the quasiparticle bands with positive and negative signs in front of the pseudomagnetic field in Eq.\ (\ref{eq.Eq.3}). Figure \ref{fig.DOS} shows the DOS for various values of $g$.

If $g$ is an even integer the DOS $D(E)$ is an analytic function of energy. However, even integers larger than $2$ are unlikely to occur since they would require the gap to open with a fractional exponent smaller than unity in at least one direction. The case $g=2$ is most naturally realized by $m=n=1$, which corresponds to a point node with linear dispersion in all directions. We will get back to this case below.

All other cases with integer $m$ and $n$ yield $g<2$ so that the DOS is not analytic at $E = \pm h$. However, the DOS can always be expanded into power series for $E<h$ and for $E>h$ separately, giving
\begin{equation}
D(E) = c_{m,n}\, h^g \sum_{j=0}^\infty \binom{g}{2j} \times
  \left\{ \begin{array}{l@{}l}
    \displaystyle \left( \frac{E}{h} \right)^{\!2j} & \mbox{for } 0 \le E \le h , \\[2ex]
    \displaystyle \left( \frac{E}{h} \right)^{\!g-2j} & \mbox{for } E > h ,
  \end{array} \right.
\label{eq.DOS.3}
\end{equation}
where the binomial coefficient is defined for any real upper argument as
\begin{equation}
\binom{a}{n} = \frac{a (a-1) (a-2) \cdots (a-n+1)}{n!}
\end{equation}
for $n=1,2,\ldots$ and $\binom{a}{0}=1$. The series over $j$ terminate if and only if $g$ is a positive integer. For later convenience, we also note the derivative
\begin{align}
D'(E) &= c_{m,n}\, g\, h^{g-1} \nonumber \\
&\quad {}\times \sum_{j=0}^\infty
  \left\{ \begin{array}{l@{}l}
    \displaystyle \binom{g-1}{2j+1} \left( \frac{E}{h} \right)^{\!2j+1}
      & \mbox{for } 0 \le E \le h , \\[2ex]
    \displaystyle \binom{g-1}{2j} \left( \frac{E}{h} \right)^{\!g-2j-1}
      & \mbox{for } E > h .
  \end{array} \right.
\label{eq.DOSp.3}
\end{align}
In the limit of uninflated point nodes, we of course obtain $D(E) = c_{m,n}\, E^g$.

\subsection{Special cases}
\label{sub.special}

For the most common case of inflated point nodes with linear dispersion,
\begin{equation}
E_\mathbf{q} = \pm h \pm \sqrt{\alpha_1^2 |\Delta_0|^2\, q_1^2
  + \alpha_2^2 |\Delta_0|^2\, q_2^2 + v_F^2 q_3^2} ,
\label{eq.Eq.4}
\end{equation}
we have $m=n=1$ and $g=2$. The dispersion is linear in all three directions, but typically with different velocities. For sufficiently high symmetry of the direction of the point node, the dispersion is isotropic in the tangential directions, i.e., $\alpha_1=\alpha_2$. The DOS reads
\begin{equation}
D(E) = c_{1,1}\, (E^2 + h^2) ,
\end{equation}
with
\begin{equation}
c_{1,1} = \frac{1}{(2\pi)^3}\, \frac{8\pi}{\alpha_1\alpha_2 |\Delta_0|^2 v_F} .
\end{equation}
The pseudomagnetic field $\mathbf{h}$ simply adds a constant to the quadratic DOS found for uninflated nodes.

As the second special case, we consider quadratic dispersion in both tangential directions, as found, for example, for pairing belonging to the irrep $E_{1g}$ of $D_{6h}$ \cite{BAM18}. Furthermore, this case can be realized by fine tuning a system with linear point nodes to a topological transition \cite{MQH13}. Generalizing the previous derivation, we treat the general anisotropic dispersion
\begin{align}
E_\mathbf{q} &= \pm h \pm \big( \alpha_1^2 |\Delta_0|^2\, q_1^4
  + \alpha_2^2 |\Delta_0|^2\, q_2^4
  + 2 \alpha_{12}^2 |\Delta_0|^2\, q_1^2 q_2^2 \nonumber \\
&\quad {}+ v_F^2 q_3^2 \big)^{1/2} .
\label{eq.Eq.5}
\end{align}
The evaluation of the DOS is relegated to Appendix \ref{app.DOS.Weyl}. The result is
\begin{equation}
D(E) = \frac{\pi\sqrt{2}}{(2\pi)^3}\,
  \frac{1}{\sqrt{\alpha_1 \alpha_2}\, |\Delta_0|\, v_F}\, G(\alpha)\,
  \big( |E + h| + |E - h| \big) ,
\label{eq.DE.6}
\end{equation}
where $\alpha \equiv \alpha_{12}^2/\alpha_1 \alpha_2$ and
\begin{align}
G(\alpha) &\equiv -\frac{1}{\sqrt{1-\alpha}} \nonumber \\
&{}\times \left[ F\left( \frac{\pi}{4} - \phi_+, \frac{2}{1-\alpha} \right)
    - F\left( \frac{\pi}{4} - \phi_-, \frac{2}{1-\alpha} \right) \right] .
\label{eq.Galpha.2}
\end{align}
Here, $F$ is the incomplete elliptic integral of the first kind and
\begin{align}
\phi_- &= \frac{\arcsin \alpha}{2} , \\
\phi_+ &= \frac{\pi - \arcsin \alpha}{2} .
\end{align}
In the limit $\alpha=0$, we have $\phi_-=0$, $\phi_+=\pi/2$, and thus
\begin{equation}
G(0) = F\left( \frac{\pi}{4}, 2 \right)
  - F\left( -\frac{\pi}{4}, 2 \right)
  = \frac{1}{2\sqrt{2\pi}}\, \Gamma^2\!\left(\frac{1}{4}\right) .
\end{equation}
Thus the DOS is
\begin{align}
D(E) &= \frac{\sqrt{\pi}}{(2\pi)^3}\,
  \frac{1}{\sqrt{\alpha_1 \alpha_2}\, |\Delta_0|\, v_F}\,
  \frac{1}{2}\, \Gamma^2\!\left(\frac{1}{4}\right) \nonumber \\
&\quad {}\times \big( |E + h| + |E - h| \big) \nonumber \\
&= \frac{c_{2,2}}{2}\, \big( |E + h| + |E - h| \big) ,
\end{align}
consistent with Eq.\ (\ref{eq.DE.g.3}) for $m=n=2$.

On the other hand, the limit $\alpha=1$ corresponds to an isotropic double-Weyl node. Here, $G(1)$ in Eq.\ (\ref{eq.Galpha.2}) is understood as
\begin{equation}
G(1) = \lim_{\alpha\to 1} G(\alpha) = \frac{\pi}{\sqrt{2}}
\end{equation}
so that
\begin{equation}
D(E) = \frac{\pi^2}{(2\pi)^3}\,
  \frac{1}{\sqrt{\alpha_1 \alpha_2}\, |\Delta_0|\, v_F}\,
  \big( |E + h| + |E - h| \big) .
\end{equation}
Clearly, the inclusion of the term $2 \alpha_{12}^2 |\Delta_0|^2\, q_1^2 q_2^2$ in the dispersion does not change the dependence on energy and pseudomagnetic field but only the prefactor.

Equation (\ref{eq.DE.6}) for general $\alpha$ can be rewritten as
\begin{equation}
D(E) = \frac{2\pi\sqrt{2}}{(2\pi)^3}\,
  \frac{1}{\sqrt{\alpha_1 \alpha_2}\, |\Delta_0|\, v_F}\, G(\alpha)
  \times \left\{\begin{array}{ll}
    h & \mbox{for } |E| \le h , \\[0.7ex]
    E & \mbox{for } |E| > h ,
  \end{array}\right.
\end{equation}
which is constant at low energies and has kinks at $E=\pm h$. We will discuss the consequences of these properties below. For uninflated double-Weyl point nodes, the DOS reduces to the absolute-value function $D(E) \propto |E|$.

Next, we turn to line nodes. For any point $\mathbf{k}_0$ on a line node, there are three orthogonal characteristic directions: the normal one, where the quasiparticle dispersion agrees with the normal state, a tangential direction parallel to the line node, where the dispersion is of course flat, and another tangential direction in which the gap opens. We denote the corresponding momentum components relative to $\mathbf{k}_0$ by $q_3$, $q_1$, and $q_2$, respectively.

We consider a circular line node of radius $k_F$, with dispersion with a general exponent $n$ in the orthogonal direction. The special case of linear dispersion appeared in the mirror plane for the $T_{2g}$ pairing state with gap amplitudes $\Delta_0 (1,i,0)$ in Refs.\ \cite{ABT17,BAM18}, for which the superconducting gap has $k_z(k_x+ik_y)$ symmetry. The quasiparticle dispersion reads
\begin{equation}
E_\mathbf{q} = \pm h \pm \sqrt{ \alpha_2^2 |\Delta_0|^2\, q_2^{2n}
  + v_F^2 q_3^2 } .
\label{eq.Eq.6}
\end{equation}
In evaluating the DOS, we have to include the $q_1$ integral along the line node, which gives a factor of $2\pi k_F$. The result is
\begin{align}
D(E) &= \frac{4\pi^{3/2}}{(2\pi)^3}\, \frac{k_F}{n\, (\alpha_2 |\Delta_0|)^{1/n}\, v_F} \nonumber \\
&\quad {}\times \big( |E + h|^{1/n} + |E - h|^{1/n} \big)\,
  \frac{\Gamma\left(\frac{1}{2n}\right)}
    {\Gamma\left(\frac{1}{2} + \frac{1}{2n}\right)} \nonumber \\
&\equiv \frac{c_{\text{line},n}}{2}\,
  \big( |E + h|^g + |E - h|^g \big) ,
\label{eq.DOS.line.4}
\end{align}
with $g=1/n$ (details are given in Appendix \ref{app.DOS.line}). For the most relevant case with linear dispersion, $n=1$ and $g=1$, we obtain the same functional form as for double-Weyl point nodes but with a different prefactor, which for line nodes is proportional to the length $2\pi k_F$ of the node.

This correspondence turns out to be more general: The case of a line node can be viewed as a special case of the general point node, where the exponent $m$ of the dispersion in one of the tangential directions is sent to infinity, leading to a flat dispersion. Equation (\ref{eq.DE.g.3}) for $m\to\infty$ has the same form as Eq.\ (\ref{eq.DOS.line.4}) for a line node but the prefactor is different since it contains the integral along the line node in the second case.

In the absence of circular symmetry, the parameters $\alpha_2$, $v_F$, and $k_F$ change along the line node but Eq.\ (\ref{eq.DOS.line.4}) should still hold with these parameters understood as averages. Such a functional form has been found for a microscopic model in Ref.~\cite{MTB19}.

Mazidian \textit{et al.}\ \cite{MQH13} have derived the DOS for superconductors with crossing uninflated line nodes with a linear or quadratic dispersion. The DOS contains logarithmic functions of energy but can be approximated by power laws with noninteger exponents $g\approx 0.8$ for crossing linear line nodes and $g\approx 0.4$ for crossing quadratic line nodes~\cite{MQH13}.

In practice, point and line nodes as well as nonequivalent point nodes can coexist. The DOS is then simply the sum of their contributions and at low energies the contributions with the smallest exponent $g$ dominate \cite{MTB19}. In the following section on observables, we therefore focus on the case of equivalent nodes. The extension to nonequivalent nodes is essentially trivial.

\section{Observables}
\label{sec.obs}

In this section, we present results for a number of observables that directly reflect the DOS and that are commonly used for the characterization of nodal superconductors. Recall that we treat the effect of interband pairing through the pseudomagnetic field $\mathbf{h}$ in effective single-band models \cite{ABT17,BAM18}. Hence, the standard derivations \cite{Sch88, Tin96} of these observables go through. While the specific heat is solely determined by the free energy and thus by the DOS, the other observables also depend on the coupling of the quasiparticles to the corresponding probe. For example, the contributions to single-particle tunneling depend on the tunneling matrix elements and the NMR relaxation rate depends on the coupling between electron and nuclear spins. In order to obtain simple closed-form results \cite{SiU91, Tin96, FeS11}, we make the usual assumption that these couplings are constant. A more quantitative treatment should not affect the functional form of the leading temperature dependence but may change the next-to-leading order.

\subsection{Single-particle tunneling}

The tunneling current between a normal conductor and the superconductor of interest provides a direct measure of the DOS of the latter if the energy dependence of the DOS of the normal conductor can be neglected. For simplicity, we also assume that the DOS of the superconductor becomes an energy-independent constant $D_0$ in its normal state. In this case, the current is given by~\cite{Tin96}
\begin{equation}
I_{sn} = \frac{G_{nn}}{e} \int_{-\infty}^\infty dE\: \frac{D(E)}{D_0}\,
  [ n_F(E) - n_F(E+eV) ] ,
\end{equation}
where $G_{nn}$ is the differential conductance if the superconductor is driven into the normal state, $n_F(E)$ is the Fermi-Dirac distribution function, and $V$ is the applied bias voltage. The differential conductance is then
\begin{equation}
G_{sn} = \frac{dI_{sn}}{dV}
  = - G_{nn} \int_{-\infty}^\infty\! dE\: \frac{D(E)}{D_0}\,
    \frac{dn_F(E+eV)}{dE} .
\end{equation}
In the zero-temperature limit, one obtains
\begin{equation}
G_{sn} = G_{nn} \int_{-\infty}^\infty\! dE\: \frac{D(E)}{D_0}\, \delta(E+eV)
  = G_{nn}\, \frac{D(-eV)}{D_0} .
\end{equation}
Thus low-temperature tunneling directly measures the superconducting DOS. In the presence of BFSs, Eq.\ (\ref{eq.DE.g.3}) implies
\begin{equation}
G_{sn} = G_{nn}\, \frac{c_{m,n}}{2D_0}\, \big( |eV + h|^g + |eV - h|^g \big) .
\end{equation}
In particular, the differential conductance is finite at zero bias, $V=0$. At non-zero temperatures, the features are smeared out over an energy scale of $k_BT$.

\subsection{Magnetic penetration depth}

The penetration depth $\lambda_i$ in the $i=x,y,z$ plane is related to the components of the response kernel $K$ by $\lambda_i \propto 1/\sqrt{K_{ii}}$ \cite{Sch88, GSE86, AGR91, ChE93, PrG06}, where the kernel $K$ links the charge current to the vector potential, $\mathbf{j} = -(e^2/mc)\, K\, \mathbf{A}$. The kernel $K$ can be decomposed into a diamagnetic part $K^d$ and a paramagnetic part $K^p$. The former is independent of temperature, while the latter satisfies~\cite{GSE86, AGR91, ChE93, PrG06}
\begin{equation}
K^d \propto \sum_n \int \frac{d^3k}{(2\pi)^3}\, \mathbf{v}_{n\mathbf{k}}
  \otimes \mathbf{v}_{n\mathbf{k}}\, \frac{dn_F}{dE} ,
\end{equation}
where $n$ is the quasiparticle-band index, $\mathbf{v}_{n\mathbf{k}} = \partial E_{n\mathbf{k}}/\partial\mathbf{k}$ is the quasiparticle velocity, and $\otimes$ denotes the Kronecker product. Introducing the DOS, the diagonal components can be written as~\cite{Ban09}
\begin{equation}
K^d_{ii} \propto \int_0^\infty dE\, D(E)\, \bigg\langle \left( \frac{\partial E}{\partial k_i}
  \right)^{\!2} \bigg\rangle_{\!E} \, \frac{dn_F}{dE} ,
\end{equation}
where $\langle \cdots \rangle_E$ is the average over all states at a given energy $E$. The averaged squared velocity can be highly anisotropic. In accord with our goal to find leading temperature dependences, we replace the average by its value at the Fermi energy, $\bar{v}_i^2$. (The averages can be evaluated explicitly. A few cases are treated in Appendix \ref{app.thermal} in the context of the thermal conductivity.) We can now write
\begin{equation}
K^d_{ii} \propto \bar{v}_i^2 \int_0^\infty dE\, D(E)\, \frac{dn_F}{dE} .
\end{equation}
At low temperature, $K^d$ is small compared to $K^p$, and we obtain
\begin{equation}
\lambda_i \propto \frac{1}{\sqrt{K^p_{ii} + K^d_{ii}}}
  \cong \frac{1}{\sqrt{K^p_{ii}}} \left( 1 - \frac{K^d_{ii}}{2 K^p_{ii}} \right)
\end{equation}
and thus
\begin{equation}
\lambda_i \cong \lambda_{i0} - \beta_i \int_0^\infty dE\, D(E)\, \frac{dn_F}{dE} ,
\label{eq.lambda.2}
\end{equation}
where $\beta_i$ is a constant proportional to $\bar{v}_i^2$. If the DOS at the Fermi energy, $D(0)$, vanishes, $\lambda_{i0}$ is the penetration depth in the limit $T\to 0$. However, for inflated nodes, the integral contains another nonvanishing term for $T\to 0$, which experimentally cannot be separated from $\lambda_{i0}$. High-precision measurements of the temperature dependence of $\lambda_i$ are insensitive to the temperature-independent part \cite{PrG06}. We therefore first consider the temperature-dependent part $\Delta\lambda_i(T) \equiv \lambda_i(T) - \lambda_i(0)$.

For a DOS of the general form of Eq.\ (\ref{eq.DE.g.3}), it is useful to split the integral at $E=h$. Substituting $u\equiv E/h$ and defining $t\equiv k_BT/h$,
integration by parts yields
\begin{align}
\lambda_i &= \lambda_{i0} - \beta_i \, \bigg[
  \left. \frac{D(uh)}{e^{u/t}+1} \right|_0^\infty
  - \int_0^1 du\, \frac{dD}{du}\, \frac{1}{e^{u/t}+1} \nonumber \\
&\quad {} - \int_1^\infty du\, \frac{dD}{du}\, \frac{1}{e^{u/t}+1} \bigg] .
\label{eq.lambda.3}
\end{align}
The first term in the angular brackets generates the new contribution to the zero-temperature penetration depth. For the other terms, we insert the series expansions in Eq.\ (\ref{eq.DOSp.3}) for $dD/du = h\, D'(E)$ and
\begin{equation}
n_F(E) = \sum_{p=1}^\infty (-1)^{p+1}\, e^{-p E/k_BT}
\label{eq.nF.3}
\end{equation}
for the Fermi-Dirac function. The result is~\cite{note.negative}
\begin{align}
\lambda_i &= \lambda_{i0} + \frac{\beta_i\, c_{m,n}\, h^g}{2}
  + \beta_i\, c_{m,n}\, g\, h^g \sum_{j=0}^\infty \sum_{p=1}^\infty 
    (-1)^{p+1} \nonumber \\
&\quad {}\times \bigg[ \int_0^1 du\, \binom{g-1}{2j+1}\, u^{2j+1}\, e^{-p u/t} \nonumber \\
&\qquad {}+ \int_1^\infty du\, \binom{g-1}{2j}\, u^{g-2j-1}\, e^{-p u/t} \bigg] \nonumber \\
&= \lambda_{i0} + \frac{\beta_i\, c_{m,n}\, h^g}{2}
  + \beta_i\, c_{m,n}\, g\, h^g \sum_{j=0}^\infty \sum_{p=1}^\infty
    (-1)^{p+1} \nonumber \\
&\quad {}\times \bigg[ \binom{g-1}{2j+1} \left(\frac{t}{p}\right)^{2j+2}
    \left[ \Gamma(2j+2) - \Gamma\left(2j+2,\frac{p}{t}\right) \right] \nonumber \\
&\qquad {}+ \binom{g-1}{2j} \left(\frac{t}{p}\right)^{g-2j}
    \Gamma\left(g-2j,\frac{p}{t}\right) \bigg] ,
\label{eq.lambda.4}
\end{align}
with the incomplete Gamma function
\begin{equation}
\Gamma(a,b) = \int_b^\infty dy\, e^{-y}\, y^{a-1} .
\end{equation}
It will prove useful to express this function as
\begin{equation}
\Gamma(a,b) \equiv e^{-b}\, b^a\, P(a,b) .
\end{equation}
In Eq.\ (\ref{eq.lambda.4}), the series over $p$ can be performed in the term only containing the complete Gamma function. Subtracting the temperature-independent part, we obtain
\begin{widetext}
\begin{align}
\Delta\lambda_i &= \beta_i\, c_{m,n}\, g\, h^g \sum_{j=0}^\infty 
  \binom{g-1}{2j+1}\, t^{2j+2}\, \Gamma(2j+2)
  \left( 1 - \frac{1}{2^{2j+1}} \right) \zeta(2j+2) \nonumber \\
&\quad {} + \beta_i\, c_{m,n}\, h^g \sum_{j=0}^\infty \sum_{p=1}^\infty
    (-1)^{p+1}\, e^{-p/t}\, \binom{g}{2j+1}
  \left[ (2j+1)\, P\left(g-2j,\frac{p}{t}\right)
    - (g-2j-1)\, P\left(2j+2,\frac{p}{t}\right) \right] ,
\label{eq.Dlambda.4}
\end{align}
\end{widetext}
where $\zeta(z)$ is the Riemann zeta function.

The function $P(a,b)$ has the series expansion
\begin{align}
P(a,b) &= \frac{1}{b} + \frac{a-1}{b^2}
  + \frac{a^2 - 3a + 2}{b^3} \nonumber \\
&\quad {} + \frac{a^3 - 6a^2 + 11a - 6}{b^4}
  + \mathcal{O}(1/b^5) .
\end{align}
For inflated nodes, $\Delta\lambda_i$ is thus generally a double series of terms proportional to $t^m\ e^{-n/t}$ with integers $m$ and $n$. At low temperatures, $k_BT \ll h$, i.e., $t\ll 1$, the exponential factors satisfy $e^{-n/t} \ll e^{-n'/t}$ for $n>n'$. In particular, a term with exponential factor is small compared to the bare power $t^m$.

The limit $h\to 0$ for uninflated nodes is not trivial since $t$ contains $h$. It is more easily calculated directly from Eq.\ (\ref{eq.lambda.2}). Integration by parts gives
\begin{align}
\lambda_i &= \lambda_{i0} - \beta_i c_{m,n} \bigg[
  E^g\, n_F(E) \big|_0^\infty
  - g \int_0^\infty dE\, E^{g-1} n_F(E) \bigg] \nonumber \\
&= \lambda_{i0} + \beta_i\, c_{m,n}\, g \nonumber \\
&\quad {}\times \left\{ \begin{array}{ll}
     k_BT\, \ln 2 & \mbox{for } g = 1, \\[0.5ex]
     \displaystyle \left(1 - \frac{1}{2^{g-1}}\right) (k_BT)^g\,
       \Gamma(g)\, \zeta(g) & \mbox{for } g \neq 1 .
  \end{array} \right.
\end{align}
The temperature-dependent part is proportional to $T^g$ for all exponents $g$~\cite{AGR91,FeS11}.

We now discuss special cases. First, if the exponent $g$ is an even integer, which is the case for linear point nodes, both sums over $j$ terminate at $j = g/2-1$. The first sum is then a polynomial in $t$ of degree $g$. In the second sum, we replace $j \to g/2-j-1$ in the terms containing $P(g-2j,p/t)$ and then find that each term in the sum cancels. Hence, the penetration depth is a polynomial in temperature of degree $g$, containing only even powers, without any exponential terms. For linear point nodes, we have $g=2$ and thus the simple result
\begin{equation}
\Delta\lambda_i = \zeta(2)\, \beta_i\, c_{1,1}\, h^2\, t^2
  = \frac{\pi^2}{6}\, \beta_i\, c_{1,1}\, (k_BT)^2 .
\end{equation}
The pseudomagnetic field only appears in the tem\-pe\-ra\-ture-in\-de\-pen\-dent part $\lambda_{i0} + \beta_i\, c_{1,1} h^2/2$. Measurements of $\Delta\lambda_i$ alone thus cannot distinguish between inflated and uninflated linear point nodes.

\begin{figure}[tb]
\centerline{\includegraphics[width=0.9\columnwidth]{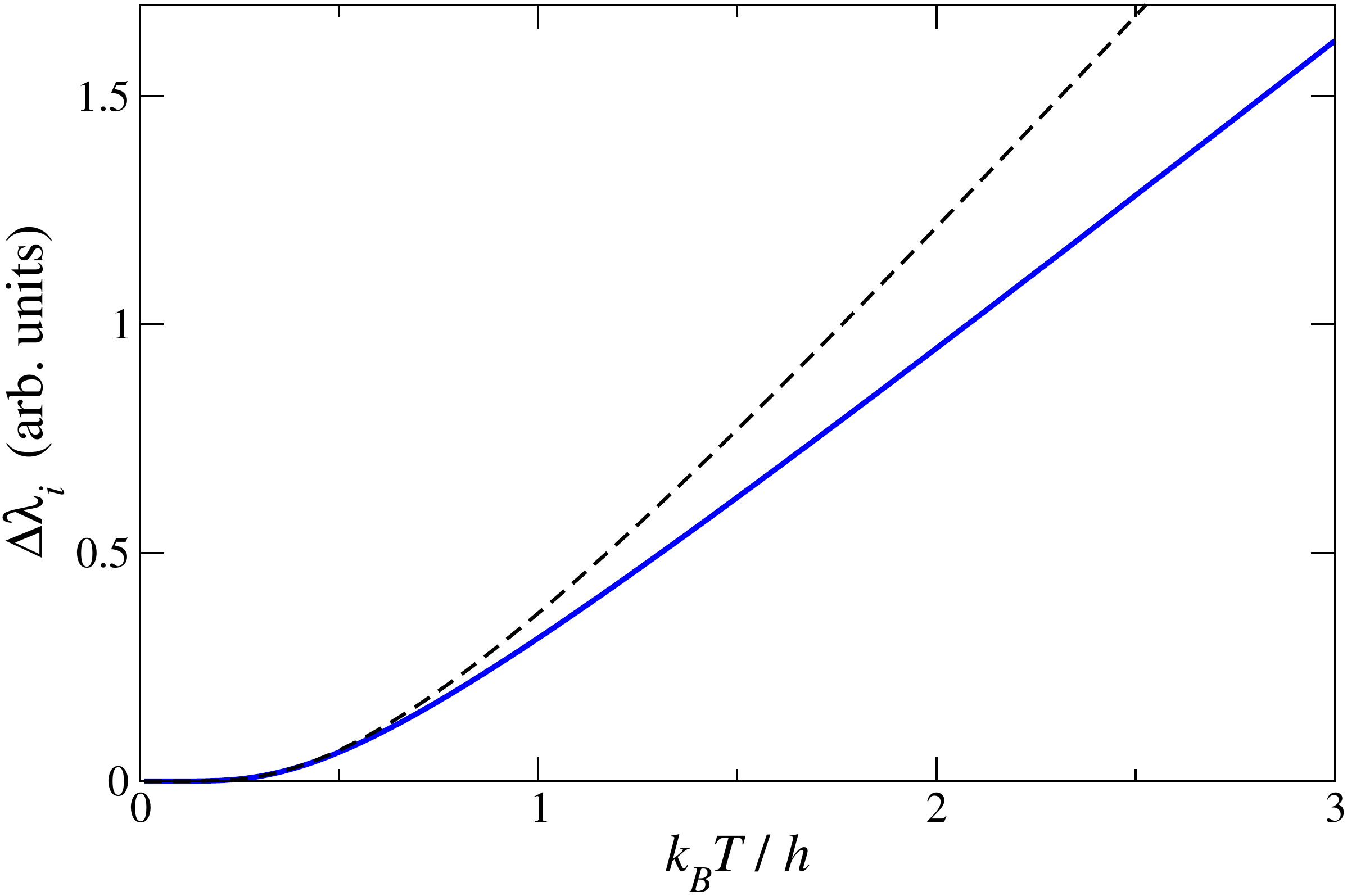}}
\caption{Temperature-dependent part $\Delta\lambda_i$ of the magnetic penetration depth (blue curve) as a function of temperature $T$ for $k_BT$ small compared to the gap amplitude $\Delta_0$. The asymptotic form for $k_BT\ll h,\Delta_0$ is shown for comparison (dashed black curve).}
\label{fig.pene_line}
\end{figure}

Second, if $g$ is an odd integer, as for double-Weyl point nodes and linear line nodes, the first sum in Eq.\ (\ref{eq.Dlambda.4}) is empty for $g=1$ and terminates at $j = (g-3)/2$ for $g\ge 3$. In the latter case, it is a polynomial in $t$ of order $g-1$, containing only even powers. The second sum terminates at $j = (g-1)/2$. The terms do not cancel, which can be inferred from the observation that $g-2j$ is odd, while $2j+2$ is even. The leading term is the one with the smallest index $p$. The most relevant case $g=1$ can be evaluated from Eq.\ (\ref{eq.lambda.3}), without series expansions,
\begin{align}
\Delta\lambda_i &= \beta_i\, c_{\text{line},1} \left[ \int_0^1 du\, \frac{0}{e^{u/t}+1}
  + \int_1^\infty du\, \frac{h}{e^{u/t}+1} \right] \nonumber \\
&= \beta_i\, c_{\text{line},1}\, k_BT\, \ln\left( 1 + e^{-h/k_BT} \right) ,
\label{eq.pene_line3}
\end{align}
written for linear line nodes. For double-Weyl point nodes, $c_{\text{line},1}$ should be replaced by $c_{2,2}$. The leading term at low temperatures is thus
\begin{equation}
\Delta\lambda_i \cong \beta_i\, c_{\text{line},1}\, k_BT\, e^{-h/k_BT} .
\end{equation}
As shown in Fig.\ \ref{fig.pene_line}, $\Delta\lambda_i$ crosses over from the exponentially suppressed form for $k_BT\ll h$ to the linear form expected for uninflated line nodes for $k_BT \gtrsim h$.

Interestingly, the exponentially suppressed temperature dependence is similar to the expression for a full gap~\cite{PrG06},
\begin{equation}
\frac{\Delta\lambda_i}{\lambda_{i0}} \approx \sqrt{\frac{\pi\Delta}{2k_BT}}\:
  e^{-\Delta/k_BT} ,
\end{equation}
where $\Delta$ is the superconducting gap assumed to be independent of momentum. However, the relevant energy scale for the case with BFSs is the pseudomagnetic field $h$, not the gap, and the temperature dependence of the prefactor is different. As noted above, the pseudomagnetic field is expected to be small compared to the gap since its leading term is of second order in the (interband) pairing \cite{ABT17,BAM18}. Nevertheless, an exponentially suppressed temperature dependence of $\Delta\lambda_i$ together with tunneling experiments showing a nodal gap would be a clear signature of inflated line and double-Weyl nodes.

Finally, if $g$ is not an integer the series in Eq.\ (\ref{eq.Dlambda.4}) do not terminate. The first series contains all even powers of $t=k_BT/h$ starting from the second. The terms in the second series all contain exponential factors of the form $e^{-p/t}$ with $p=1,2,\ldots$ and positive integer powers of $t$ [note that $P(a,b)$ consists of negative integer powers of $b=p/t$]. We note that the exponential terms found whenever $g$ is not an even integer result from the nonanalytical energy dependence of the DOS at $E=\pm h$, see Fig.\ \ref{fig.DOS}. At low temperatures, the second-order term of the first series dominates $\Delta\lambda_i$ and we obtain
\begin{equation}
\Delta\lambda_i \cong \frac{\pi^2}{12}\, \beta_i\, c_{m,n}\, g(g-1)\, h^{g-2}\, (k_BT)^2 .
\end{equation}
Hence, the exponent describing the temperature dependence does not provide information on the nature of the nodes.

So far, we have only considered the temperature-dependent part $\Delta\lambda_i$. The absolute value of $\lambda_i$ can be measured, for example, by muon spin rotation ($\mu$SR) \cite{SBK00, PrG06}. Conventionally, the penetration depth $\lambda \propto 1/\sqrt{n_s}$ is expressed in terms of the superfluid density $n_s$, and $n_s$ reaches the full electron concentration $n_e$ for $T\to 0$ if the superconducting volume fraction is $100\,\%$ (the anisotropic case requires additional analysis). $\lambda$ is of interest here since in the presence of BFSs, $n_s$ is not expected to reach $n_e$ because of the surviving quasiparticle contribution. Hence, $n_s/n_e$ smaller than the superconducting volume fraction would be a strong signature. However, careful modeling of the $\mu$SR signal would be needed since we are dealing with TRS-breaking superconductors, for which the contribution from magnetic penetration is difficult to disentangle from the intrinsic magnetic field.

\subsection{Specific heat}

We next address the electronic contribution to the specific heat, which is given by \cite{SiU91, FeS11, MQH13}
\begin{equation}
c = -\frac{1}{T} \int_0^\infty dE\, D(E)\, E^2\, \frac{dn_F}{dE} .
\label{eq.c.2}
\end{equation}
It is again useful to split the integral at $E=h$ and to substitute $u=E/h$.
Integration by parts yields
\begin{align}
c &= -\frac{h^2}{T}\, \bigg[
  \left. \frac{u^2 D(uh)}{e^{u/t}+1} \right|_0^\infty
  - \int_0^1 du\, \frac{2u D + u^2\, \frac{dD}{du}}{e^{u/t}+1}
    \nonumber \\
&\quad {}- \int_1^\infty du\, \frac{2u D + u^2\, \frac{dD}{du}}{e^{u/t}+1} \bigg] ,
\label{eq.c.4}
\end{align}
where $t = k_BT/h$.

For now, we assume that the temperature dependence of the DOS due to the pairing amplitude $\Delta_0(T)$ is negligible. Inserting the expansions in Eqs.\ (\ref{eq.DOS.3}), (\ref{eq.DOSp.3}), and (\ref{eq.nF.3}), we obtain
\begin{widetext}
\begin{align}
c
&= \frac{2}{T}\, c_{m,n}\, h^{g+2}
  \sum_{j=0}^\infty \sum_{p=1}^\infty (-1)^{p+1}\,
  \bigg[ \binom{g}{2j} \left(\frac{t}{p}\right)^{2j+2}
    \left[ \Gamma(2j+2) - \Gamma\left(2j+2,\frac{p}{t}\right) \right]
  + \binom{g}{2j} \left(\frac{t}{p}\right)^{g-2j+2}
    \Gamma\left(g-2j+2,\frac{p}{t}\right) \bigg] \nonumber \\
&\quad {}+ \frac{1}{T}\, c_{m,n}\, g\, h^{g+2}
  \sum_{j=0}^\infty \sum_{p=1}^\infty (-1)^{p+1}\,
  \bigg[ \binom{g-1}{2j+1} \left(\frac{t}{p}\right)^{2j+4}
    \left[ \Gamma(2j+4) - \Gamma\left(2j+4,\frac{p}{t}\right) \right] \nonumber \\
&\quad {}+ \binom{g-1}{2j} \left(\frac{t}{p}\right)^{g-2j+2}
    \Gamma\left(g-2j+2,\frac{p}{t}\right) \bigg] .
\end{align}
The series over $p$ can be performed in the terms only containing the complete Gamma function. After additional simplifications, we obtain
\begin{align}
c
&= 2k_B\, c_{m,n}\, h^{g+1}
   \sum_{j=0}^\infty \binom{g}{2j}\, (1+j)\, t^{2j+1}\, \Gamma(2j+2)
   \left( 1 - \frac{1}{2^{2j+1}} \right) \zeta(2j+2) \nonumber \\
&\quad{} + \frac{1}{T}\, c_{m,n}\, h^{g+2}
  \sum_{j=0}^\infty \sum_{p=1}^\infty (-1)^{p+1}\, e^{-p/t}\, \binom{g}{2j}
  \left[ (g-2j+2)\, P\left( g-2j+2, \frac{p}{t} \right)
  - (2j+2)\, P\left( 2j+2, \frac{p}{t} \right) \right] .
\label{eq.c.6}
\end{align}
\end{widetext}
For uninflated nodes, integration by parts in Eq.\ (\ref{eq.c.2}) gives
\begin{align}
c &= -\frac{c_{m,n}}{T} \bigg[ E^{g+2}\, n_F(E) \big|_0^\infty \nonumber \\
&\quad {}- (g+2) \int_0^\infty dE\, E^{g+1}\, n_F(E) \bigg] \nonumber \\
&= k_B\, c_{m,n}\, (g+2)
  \left( 1 - \frac{1}{2^{g+1}} \right) (k_BT)^{g+1} \nonumber \\
&\quad {}\times \Gamma(g+2)\, \zeta(g+2) .
\end{align}
Hence, the electronic contribution to the specific heat is proportional to $T^2$ for linear line nodes and to $T^3$ for linear point nodes, as is well known~\cite{SiU91, MQH13}.

The discussion of special cases with BFSs follows the one for the penetration depth and we can be brief. For even integer $g$, the specific heat is a polynomial in temperature of order $g+1$, containing only odd powers. In particular, for inflated linear point nodes ($g=2$), we obtain
\begin{equation}
c
  = k_B c_{m,n} \left[ \frac{\pi^2}{6}\, h^2\, k_BT + \frac{7\pi^4}{30}\, (k_BT)^3 \right]
\label{eq.c.8}
\end{equation}
so that the Sommerfeld coefficient is
\begin{equation}
\gamma = \frac{c}{T}
  = k_B^2 c_{m,n} \left[ \frac{\pi^2}{6}\, h^2 + \frac{7\pi^4}{30}\, (k_BT)^2 \right] .
\end{equation}
Since the DOS is finite at zero energy we obtain a finite Sommerfeld coefficient for $T\to 0$.

There is experimental evidence for a finite Sommerfeld coefficient in a number of compounds, e.g., thorium-doped $\mathrm{UBe}_{13}$ \cite{ZDS04}, $\mathrm{UPt}_3$ \cite{BKL94,JoT02}, $\mathrm{URu_2Si_2}$ (from thermal-conductivity measurements) \cite{KIS07}, and $\mathrm{UTe}_2$ \cite{RED19,ANH19}. A residual density of states and an associated nonzero Sommerfeld coefficient can more conventionally be caused by disorder. However, in particular for $\mathrm{UTe}_2$, the small variation of $T_c$ between various samples and the large and sharp jump in the specific heat at $T_c$ suggest that disorder is weak \cite{RED19}. Then BFSs are a natural explanation, as also noted in Ref.\ \cite{ShP19}. In addition, the observed scaling $c(T)-c(0) \sim T^{3.2}$, and thus $\gamma(T)-\gamma(0) \sim T^{2.2}$, agrees reasonably well with the exponent of $2$ expected for inflated point nodes. On the other hand, more recent specific-heat measurements for $\mathrm{UTe}_2$ by Metz \textit{et al.}\ \cite{MBR19} are consistent with uninflated point nodes (and thus no residual Sommerfeld coefficient from the superconducting state) plus a diverging contribution proportional to $T^{-0.35}$ from quantum-critical magnetic fluctuations. The authors also observe the scaling $\Delta\lambda \sim T^2$ for the magnetic penetration depth \cite{MBR19}, which, however, does not discriminate between uninflated and inflated point nodes, as discussed above. Moreover, they find a $T^3$ power law for the electronic contribution to the thermal conductivity, which is consistent with uninflated point nodes \cite{MBR19}, as shown below.

For odd integer $g$, both $j$ series in Eq.\ (\ref{eq.c.6}) terminate. The first is a polynomial in $t$ of order $g$, containing only odd powers. The terms in the second series do not cancel, giving exponentially suppressed corrections to the polynomial. For $g=1$ (linear line nodes and double-Weyl point nodes), the two leading terms read
\begin{equation}
c
\cong k_B\, c_{\text{line},1} \left[ \frac{\pi^2}{6}\, h\, k_BT
  + h^2\, e^{-h/k_BT} \right]
\label{eq.c.10}
\end{equation}
so that
\begin{equation}
\gamma \cong k_B^2\, c_{\text{line},1} \left[ \frac{\pi^2}{6}\, h
  + \frac{h^2}{k_BT}\, e^{-h/k_BT} \right],
\end{equation}
written for linear line nodes. For double-Weyl point nodes, $c_{\text{line},1}$ should be replaced by $c_{2,2}$. So unlike for the temperature-dependent part of the penetration depth, the leading term in $c$ and $\gamma$ is not exponentially suppressed but is given by the contribution from the residual DOS, while the first correction is exponential. Both leading terms vanish for $h\to 0$ so that the approach to the result $\gamma \propto T$ for uninflated nodes is not trivial.

If $g$ is not an integer, the series in Eq.\ (\ref{eq.c.6}) do not terminate. The two leading terms are
\begin{equation}
c \cong
  k_B\, c_{m,n}\, h^g \left[ \frac{\pi^2}{6}\, k_BT
  + \frac{7\pi^4}{30}\, g\, \frac{(k_BT)^3}{h^2} \right]
\end{equation}
and thus
\begin{equation}
\gamma \cong k_B^2\, c_{m,n}\, h^g \left[ \frac{\pi^2}{6}
  + \frac{7\pi^4}{30}\, g\, \frac{(k_BT)^2}{h^2} \right] .
\end{equation}
Considering all cases together, the first correction to the residual Sommerfeld coefficient thus scales as $T^2$ unless $g$ equals unity.

Let us now consider the case that the DOS depends on temperature through the pairing amplitude $\Delta_0(T)$. Then two additional contributions to the specific heat appear. On the one hand, the mean-field decoupling of the interaction
\begin{align}
H_\mathrm{int} &= \frac{1}{2} \sum_{\mathbf{k}\mathbf{k}'} \sum_{\sigma_1\sigma_2\sigma_3\sigma_4}
  V_{\sigma_1\sigma_2\sigma_3\sigma_4}(\mathbf{k},\mathbf{k}') \nonumber \\
&\quad{} \times c^\dagger_{-\mathbf{k},\sigma_1} c^\dagger_{\mathbf{k},\sigma_2}
  c^{\phantom{\dagger}}_{\mathbf{k}',\sigma_3} c^{\phantom{\dagger}}_{-\mathbf{k}',\sigma_4} ,
\end{align}
where $c^\dagger_{\mathbf{k},\sigma}$ and $c^{\phantom{\dagger}}_{\mathbf{k},\sigma}$ are electron creation and annihilation operators, respectively, produces a constant energy shift
\begin{align}
\Delta E_\mathrm{mf} &= - \frac{1}{2} \sum_{\mathbf{k}\mathbf{k}'} \sum_{\sigma_1\sigma_2\sigma_3\sigma_4}
  V_{\sigma_1\sigma_2\sigma_3\sigma_4}(\mathbf{k},\mathbf{k}') \nonumber \\
&\quad{} \times \big\langle c^\dagger_{-\mathbf{k},\sigma_1}
    c^\dagger_{\mathbf{k},\sigma_2} \big\rangle\,
  \big\langle c^{\phantom{\dagger}}_{\mathbf{k}',\sigma_3}
    c^{\phantom{\dagger}}_{-\mathbf{k}',\sigma_4} \big\rangle ,
\end{align}
which is proportional to $|\Delta_0|^2$. This term leads to a correction to the specific heat of the form
\begin{equation}
\Delta c_\mathrm{mf} \propto |\Delta_0|\, \frac{d|\Delta_0|}{dT} .
\end{equation}
On the other hand, the quasiparticle contribution to the specific heat, given in Eq.\ (\ref{eq.c.2}), obtains an additional term since it is the temperature derivative of the internal-energy density, which now depends on temperature also through $\Delta_0$,
\begin{equation}
c = \int_0^\infty dE
  \left[ -D(E)\, \frac{E^2}{T}\, \frac{dn_F}{dE}
  + \frac{dD}{d|\Delta_0|}\, \frac{d|\Delta_0|}{dT}\, E\, n_F(E) \right] .
\end{equation}
The new, second term can be evaluated by series expansion, similar to the leading term considered above, which we do not show here. In any case, we observe that both new terms are proportional to the derivative $d|\Delta_0|/dT$. Evaluation of this quantity requires, in the weak-coupling case, the solution of a BCS gap equation for a microscopic model, which lies outside of the scope of this paper.
Note that additional terms due to the temperature dependence of the pairing amplitude occur only in the specific heat and the related thermal conductivity but not in the other observables considered here.

\subsection{Spin-lattice relaxation rate}

The NMR spin-lattice relaxation rate is given by~\cite{HeS59, Sch88, SiU91, FeS11}
\begin{equation}
\frac{1}{T_1T} = -\beta_\mathrm{NMR} \int_0^\infty dE\, D^2(E)\, \frac{dn_F}{dE} ,
\label{eq.T1.2}
\end{equation}
where $\beta_\mathrm{NMR}$ is a constant. The general expression \cite{HeS59, Sch88, SiU91} contains a coherence factor, which involves constant-energy averages of the gap amplitude. We assume an unconventional pairing state belonging to a nontrivial irreducible representation of the point group \cite{note.irrep}. Under this condition, these averages vanish \cite{SiU91}. In addition, we have neglected the small nuclear resonance frequency, which is reasonable for nodal states \cite{SiU91}. By splitting the integral at $E=h$, substituting $u=E/h$, and integrating by parts, we obtain
\begin{align}
\frac{1}{T_1T} &= -\beta_\mathrm{NMR}\, \bigg[
  \left. \frac{D^2(uh)}{e^{u/t}+1} \right|_0^\infty
  - \int_0^1 du\, \frac{2D\, \frac{dD}{du}}{e^{u/t}+1} \nonumber \\
&\quad {}- \int_1^\infty du\, \frac{2D\, \frac{dD}{du}}{e^{u/t}+1} \bigg] .
\end{align}
Inserting the expansions in Eqs.\ (\ref{eq.DOS.3}), (\ref{eq.DOSp.3}), and (\ref{eq.nF.3}), we obtain
\begin{widetext}
\begin{align}
\frac{1}{T_1T} &= \frac{\beta_\mathrm{NMR}\, c_{m,n}^2\, h^{2g}}{2}
  + 2 \beta_\mathrm{NMR}\, c_{m,n}^2\, g\, h^{2g}
    \sum_{j=0}^\infty \sum_{k=0}^\infty \sum_{p=1}^\infty (-1)^{p+1} \nonumber \\
&\quad {}\times \bigg[ \int_0^1 du\, \binom{g}{2j} \binom{g-1}{2k+1}\, u^{2j+2k+1}\, e^{-pu/t}
  + \int_1^\infty du\, \binom{g}{2j} \binom{g-1}{2k}\, u^{2g-2j-2k-1}\, e^{-pu/t} \bigg] \nonumber \\
&= \frac{\beta_\mathrm{NMR}\, c_{m,n}^2\, h^{2g}}{2} \nonumber \\
&\quad {}+ 2 \beta_\mathrm{NMR}\, c_{m,n}^2\, g\, h^{2g}
  \sum_{j=0}^\infty \sum_{k=0}^\infty \binom{g}{2j} \binom{g-1}{2k+1}\,
    t^{2j+2k+2}\, \Gamma(2j+2k+2) \left( 1 - \frac{1}{2^{2j+2k+1}} \right) \zeta(2j+2k+2) \nonumber \\
&\quad {}+ 2 \beta_\mathrm{NMR}\, c_{m,n}^2\, h^{2g}
  \sum_{j=0}^\infty \sum_{k=0}^\infty \sum_{p=1}^\infty (-1)^{p+1}\, e^{-p/t}\,
    \binom{g}{2j} \binom{g}{2k+1} \nonumber \\
&\quad {}\times \bigg[ (2k+1)\, P\left( 2g-2j-2k, \frac{p}{t} \right)
  - (g-2k-1)\, P\left( 2j+2k+2, \frac{p}{t} \right) \bigg] .
\label{eq.T1.5}
\end{align}
For uninflated nodes, integration by parts in Eq.\ (\ref{eq.T1.2}) yields
\begin{align}
\frac{1}{T_1T} &= -\beta_\mathrm{NMR}\, c_{m,n}^2\, \bigg[ E^{2g}\, n_F(E) \big|_0^\infty
  - 2g \int_0^\infty dE\, E^{2g-1}\, n_F(E) \bigg] \nonumber \\
&= 2\beta_\mathrm{NMR}\, c_{m,n}^2\, g
  \times \left\{ \begin{array}{ll}
    \displaystyle k_BT\, \ln 2 &
      \displaystyle \mbox{for } g = \frac{1}{2} , \\[1.8ex]
    \displaystyle \left( 1 - \frac{1}{2^{2g-1}} \right)
      (k_BT)^{2g}\, \Gamma(2g)\, \zeta(2g) &
      \displaystyle \mbox{for } g \neq \frac{1}{2} .
  \end{array} \right.
\label{eq.T1.uninfl.2}
\end{align}
\end{widetext}
The spin-lattice relaxation rate is proportional to $T^2$ for linear line nodes and to $T^4$ for linear point nodes~\cite{SiU91}.

Returning to inflated nodes, we note that there is always a residual relaxation rate at zero temperature due to the nonzero DOS at the Fermi energy. The form of the leading correction depends on the exponent $g$. For even integer $g$, the first double series in $j$ and $k$ in Eq.\ (\ref{eq.T1.5}) terminates and the second vanishes, giving a polynomial in $t$ of order $2g$, containing only even powers. For $g=2$, we find
\begin{equation}
\frac{1}{T_1T}
\cong \beta_\mathrm{NMR}\, c_{m,n}^2
  \left[ \frac{1}{2}\, h^4 + \frac{\pi^2}{3}\, h^2\, (k_BT)^2 \right]
\end{equation}
for the two leading terms. Note that both terms vanish for $h\to 0$. Only the following term, which is proportional to $h^0\, T^4$, survives and gives the result (\ref{eq.T1.uninfl.2}) for uninflated nodes.

For odd integer $g$, the $j$, $k$ series in Eq.\ (\ref{eq.T1.5}) both terminate. The first vanishes for $g=1$ and is a polynomial in $t$ of order $2g-2$ for $g>1$, containing only even powers. The second gives powers multiplied by exponentials. For $g=1$, the two leading terms read
\begin{equation}
\frac{1}{T_1T} \cong
\beta_\mathrm{NMR}\, c_{\text{line},1}^2
  \left[ \frac{1}{2}\, h^2 + 2 h\, k_BT\, e^{-h/k_BT} \right] ,
\end{equation}
written for linear line nodes. Similar to the Sommerfeld coefficient, both leading terms vanish for $h\to 0$ so that the approach to the result $1/T_1T \propto T^2$ for uninflated nodes is not trivial.

If $g$ is not an integer, the series in Eq.\ (\ref{eq.T1.5}) do not terminate. The two leading terms are
\begin{equation}
\frac{1}{T_1T} \cong
\beta_\mathrm{NMR}\, c_{m,n}^2 \left[ \frac{1}{2}\, h^{2g}
  + \frac{\pi^2}{6}\, g(g-1)\, h^{2g-2} (k_BT)^2 \right] .
\end{equation}
It is worth pointing out that the spin-lattice relaxation rate depends on the square of the DOS. Since we expect the DOS due to the BFSs to be small the resulting terms in $1/T_1T$ should also be small, which implies that they require very low temperatures in order to be observed.

\subsection{Thermal conductivity}

The electronic contribution to the thermal conductivity can be described in semiclassical Boltzmann theory, as derived by Bardeen, Rickayzen, and Tewordt \cite{BRT59}, starting from BCS theory. This derivation does not rely on the underlying mechanism but only on the presence of Bogoliubov quasiparticles as low-energy excitations. (The condensate itself does not carry entropy and thus does not contribute to thermal transport.) Boltzmann theory is limited to low energies and long wavelengths. These conditions are satisfied since we consider the uniform response to elastic scattering, which is justified at low temperatures, where the contribution from electron-phonon scattering freezes out.

In addition, we employ a relaxation-time approximation with a constant relaxation time $\tau$. The relaxation-time form of the scattering integral becomes exact for elastic point scattering but the resulting relaxation time generally depends on energy. Our actual approximation thus consists of taking the same value of $\tau$ for all low-energy quasiparticle states, which is analogous to the assumption of constant couplings for the previously discussed observables.

Hence, we employ Boltzmann theory in the re\-la\-xa\-tion-time approximation and write the energy current density as~\cite{Mar10, Ket16}
\begin{equation}
\mathbf{j}_E(\mathbf{r}) = \tau \sum_n \int \frac{d^3k}{(2\pi)^3}\,
  E^2_{n\mathbf{k}}\, \mathbf{v}_{n\mathbf{k}}\,
  \mathbf{v}_{n\mathbf{k}} \cdot \frac{\mbox{\boldmath$\nabla$} T(\mathbf{r})}{T(\mathbf{r})}\,
  \frac{dn_F}{dE} ,
\end{equation}
where $\tau$ is the relaxation time, $n$ is the quasiparticle-band index, and $\mathbf{v}_{n\mathbf{k}} = \partial E_{n\mathbf{k}}/\partial \mathbf{k}$. Taking $\mbox{\boldmath$\nabla$} T/T$ to be weakly varying in space, we can write
\begin{equation}
\mathbf{j}_E(\mathbf{r}) = -\kappa\, \mbox{\boldmath$\nabla$} T ,
\end{equation}
with the thermal conductivity tensor
\begin{equation}
\kappa
= - \frac{\tau}{T} \int_0^\infty dE\, D(E)\, E^2
  \left\langle \frac{\partial E}{\partial \mathbf{k}}
    \otimes \frac{\partial E}{\partial \mathbf{k}} \right\rangle_{\!E} \, \frac{dn_F}{dE} .
\label{eq.kappa.2}
\end{equation}
The last expression is similar to Eq.\ (\ref{eq.c.2}) for the specific heat but is a tensor due to the velocity product. The tensorial character is crucial when summing contributions from multiple nodes.

We take $D(E)$ to be unaffected by disorder, which is valid if the energy scale $\gamma$ characteristic for the smearing of the DOS by disorder is small compared to all other energy scales, i.e, in the clean limit. The scale $\gamma$ is on the order of the relaxation rate $1/\tau$ so that our results are valid for $1/\tau \ll k_BT$. Thus, we do not address the \emph{universal limit} \cite{GYS96, DuL00}, which applies for $k_BT \lesssim 1/\tau$. In this regime, the low-energy DOS becomes proportional to $\gamma\sim 1/\tau$, which cancels the prefactor in Eq.\ (\ref{eq.kappa.2}) and makes the thermal conductivity independent of $\tau$.

The evaluation of the average for the general dispersions considered so far is cumbersome and we restrict ourselves to linear point and line nodes. For inflated point nodes with the linear dispersion given in Eq.\ (\ref{eq.Eq.4}), we find the thermal-conductivity tensor
\begin{equation}
\kappa = c\tau \left( \begin{array}{ccc}
    \displaystyle \frac{\alpha_1^2 |\Delta_0|^2}{3} & 0 & 0 \\
    0 & \displaystyle \frac{\alpha_2^2 |\Delta_0|^2}{3} & 0 \\
    0 & 0 & \displaystyle \frac{v_F^2}{3}
  \end{array} \right) ,
\end{equation}
where $c$ is the specific heat, which for linear point nodes has the form in Eq.\ (\ref{eq.c.8}). The derivation is sketched in Appendix \ref{app.thermal.point}. The result is rather simple since the average $\langle \partial E/\partial \mathbf{k}\, \otimes\, \partial E/\partial \mathbf{k} \rangle_E$ is independent of energy in this case. The temperature dependence of $\kappa$ is then due to the one of the specific heat. The thermal conductivity is diagonal in the chosen basis but typically much larger in the direction orthogonal to the normal-state Fermi surface than tangential to it since $v_F \gg \alpha_{1,2} |\Delta_0|$. However, for high-symmetry states, the anisotropy will disappear when the contributions from all nodes are summed up.

For the case of an inflated linear line node, we again consider a circular line of radius $k_F$, which we take to lie in the $k_xk_y$ plane. The dispersion reads
\begin{equation}
E_\mathbf{k} = \pm h
  \pm \sqrt{ v_F^2\, (k_\rho-k_F)^2 + \alpha_2^2 |\Delta_0|^2\, k_z^2 } ,
\end{equation}
where $k_\rho=(k_x^2+k_y^2)^{1/2}$ and $k_z$ are cylindrical coordinates describing the momentum $\mathbf{k}$. We find
\begin{equation}
\kappa = c\tau \left( \begin{array}{ccc}
    \displaystyle \frac{v_F^2}{4} & 0 & 0 \\
    0 & \displaystyle \frac{v_F^2}{4} & 0 \\
    0 & 0 & \displaystyle \frac{\alpha_2^2 |\Delta_0|^2}{2}
  \end{array} \right) ,
\end{equation}
where the specific heat $c$ for linear line nodes is given in Eq.\ (\ref{eq.c.10}). Details are given in Appendix \ref{app.thermal.line}. The result is again simple since $\langle \partial E/\partial \mathbf{k}\, \otimes\, \partial E/\partial \mathbf{k} \rangle_E$ is energy independent. The thermal conductivity is diagonal and generally highly anisotropic also in this case.
The thermal conductivity of the $T_{2g}$ pairing state with gap amplitudes $\Delta_0(1,i,0)$ \cite{ABT17,BAM18} and
$k_z(k_x+ik_y)$ symmetry is expected to be dominated by the inflated line node since it contributes much more to the DOS than the inflated point nodes. Consequently, such a pairing state would lead to a much larger thermal conductivity in the plane of the line node than in the perpendicular direction, while both components would scale linearly with temperature and deviations from linearity would be exponentially small.

For all other cases of inflated nodes, including double-Weyl point nodes, the average $\langle \partial E/\partial \mathbf{k}\, \otimes\, \partial E/\partial \mathbf{k} \rangle_E$ depends on energy and thus does not come out of the integral in Eq.\ (\ref{eq.kappa.2}). The thermal conductivity is then not simply proportional to the specific heat.

\section{Search for surface states}
\label{sec.surface}

As noted in Sec.\ \ref{sec.intro}, nontrivial topological properties in the bulk are often revealed by unconventional surface states. These states contribute to transport and tunneling and can, in principle, be observed by ARPES. For example, line nodes in noncentrosymmetric superconductors lead to the appearance of flat zero-energy surface bands in the two-dimensional Brillouin zone of the surface, which are bounded by the projections of the line nodes into this Brillouin zone \cite{TMY10,ScR11,BST11,SBT12,ScB15,TSA17}. The BFSs are protected by a $\mathbb{Z}_2$ invariant \cite{KST14,ZSW16,ABT17,BAM18}. In this section, we address the question of whether they are accompanied by surface states, in particular in the momentum-space region bounded by the projection of the Fermi surfaces. To be able to study this case, we set up a model with \emph{open}~BFSs.

One might not expect surface states of topological origin for the following reason: The $\mathbb{Z}_2$ invariant has been identified as the relative sign of a Pfaffian of a unitarily transformed Bogoliubov--de Gennes Hamiltonian \cite{ABT17,BAM18}. The derivation shows that the global sign of this Pfaffian depends on the unitary transformation. Hence, it does not make sense to speak of the momentum-space region enclosed by the BFSs as topologically nontrivial and the region outside as topologically trivial. There is thus no clear reason to expect surface bands either inside or outside of the projection of the BFSs. On the other hand, we can also find an argument as to why surface states could exist: A self-consistent real-space calculation would show that the pairing amplitude $\Delta_0(z)$ changes with the distance $z$ from the surface. Now consider the case that this change is very slow so that one can infer the properties at a depth $z$ by assuming a uniform pairing amplitude with the value $\Delta_0(z)$. Assume further that the pairing amplitude approaches zero at the surface. Then the BFSs vary as a function of $z$ and in particular shrink to point or line nodes for $z$ approaching the surface. Hence, for any momentum inside a bulk BFS, there is some depth for which the BFS crosses that point so that there is a zero-energy state bound to the surface.

\begin{figure}[tb]
\centerline{\includegraphics[width=0.8\columnwidth]{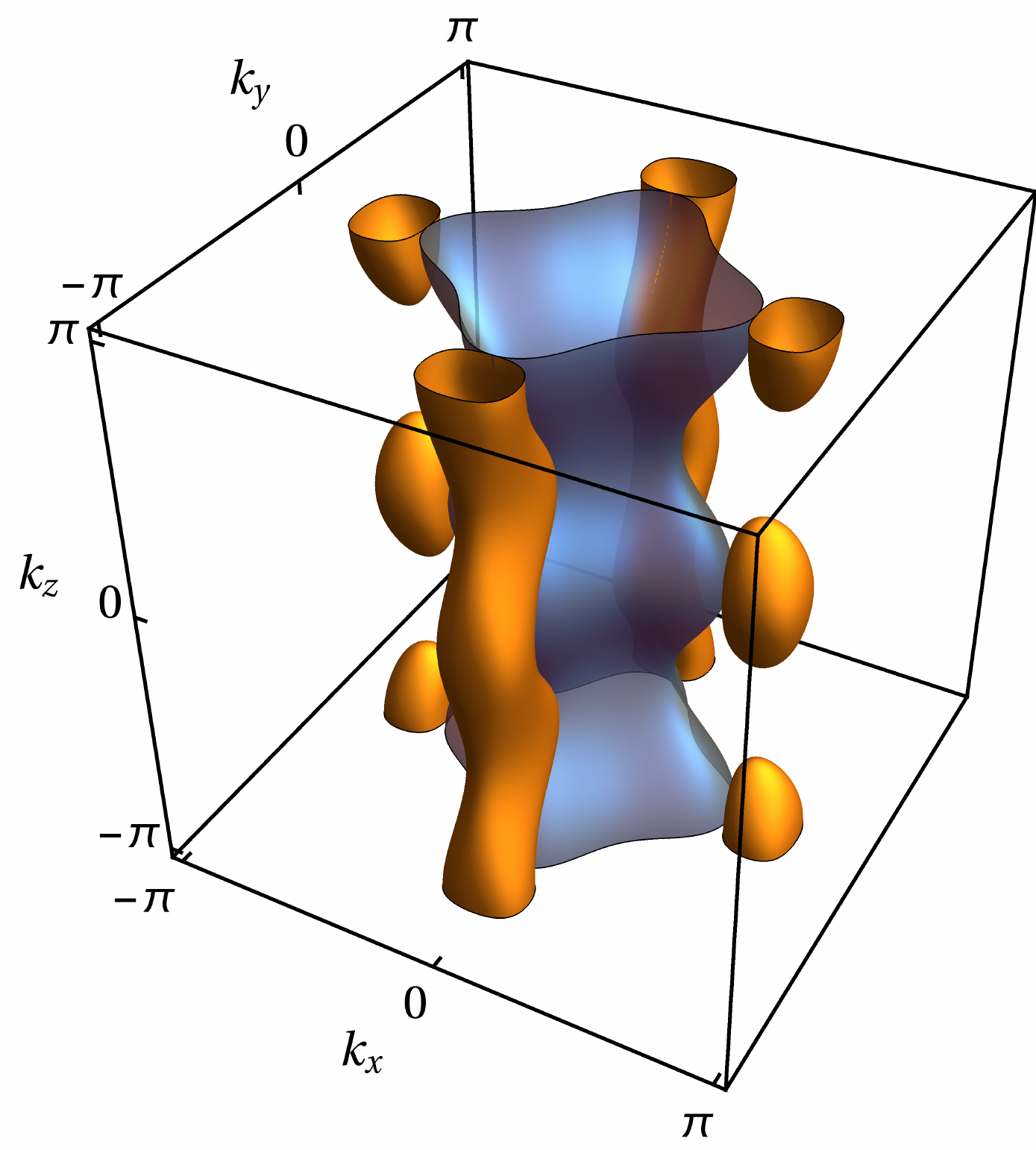}}
\caption{Bulk Fermi surfaces in the normal (semitransparent gray) and superconducting (orange) states of the model used in the search for surface states. Due to the weak dispersion in the \textit{z} direction, the normal-state Fermi surface and two of the BFSs are deformed open cylinders.}
\label{fig.FS}
\end{figure}

In order to construct a model with open BFSs, we start from the model for superconductivity of angular-momentum $J=3/2$ fermions studied in Ref.\ \cite{ABT17} and deform the normal-state Hamiltonian in such a way that the hopping in the \textit{z} direction becomes weak. The superconducting state is characterized by the uniform pairing potential (the off-diagonal block in the Bogoliubov--de Gennes Hamiltonian)
\begin{equation}
\Delta = \Delta_0\, ( \eta_{xy} + i\eta_{xz} ) ,
\end{equation}
with \cite{ABT17}
\begin{align}
\eta_{xy} &= \frac{J_x J_y + J_y J_x}{\sqrt{3}}\, U_T , \\
\eta_{xz} &= \frac{J_x J_z + J_z J_x}{\sqrt{3}}\, U_T ,
\end{align}
where $J_\alpha$ are standard $4\times 4$ angular-momentum $J=3/2$ matrices and
\begin{equation}
U_T = \left(\begin{array}{cccc}
  0 & 0 & 0 & 1 \\
  0 & 0 & -1 & 0 \\
  0 & 1 & 0 & 0 \\
  -1 & 0 & 0 & 0
  \end{array}\right)
\end{equation}
is the unitary part of the time-reversal operator. This pairing state is rotated relative to the $(1,i,0)$ state in the $T_{2g}$ irrep in Ref.\ \cite{ABT17} so that the inflated line node lies in the $k_yk_z$ plane. We emphasize that the details of the model do not matter; if BFSs led to protected surface states they would do so for any model. Hence, we do not write down the specific model but only show the resulting normal-state and Bogoliubov Fermi surfaces in Fig.\ \ref{fig.FS}. The inflated line node in the $k_yk_z$ plane is not a torus in this case but is split into two open tubes along the $k_z$ direction.

\begin{figure}[tb]
\centerline{\includegraphics[width=\columnwidth]{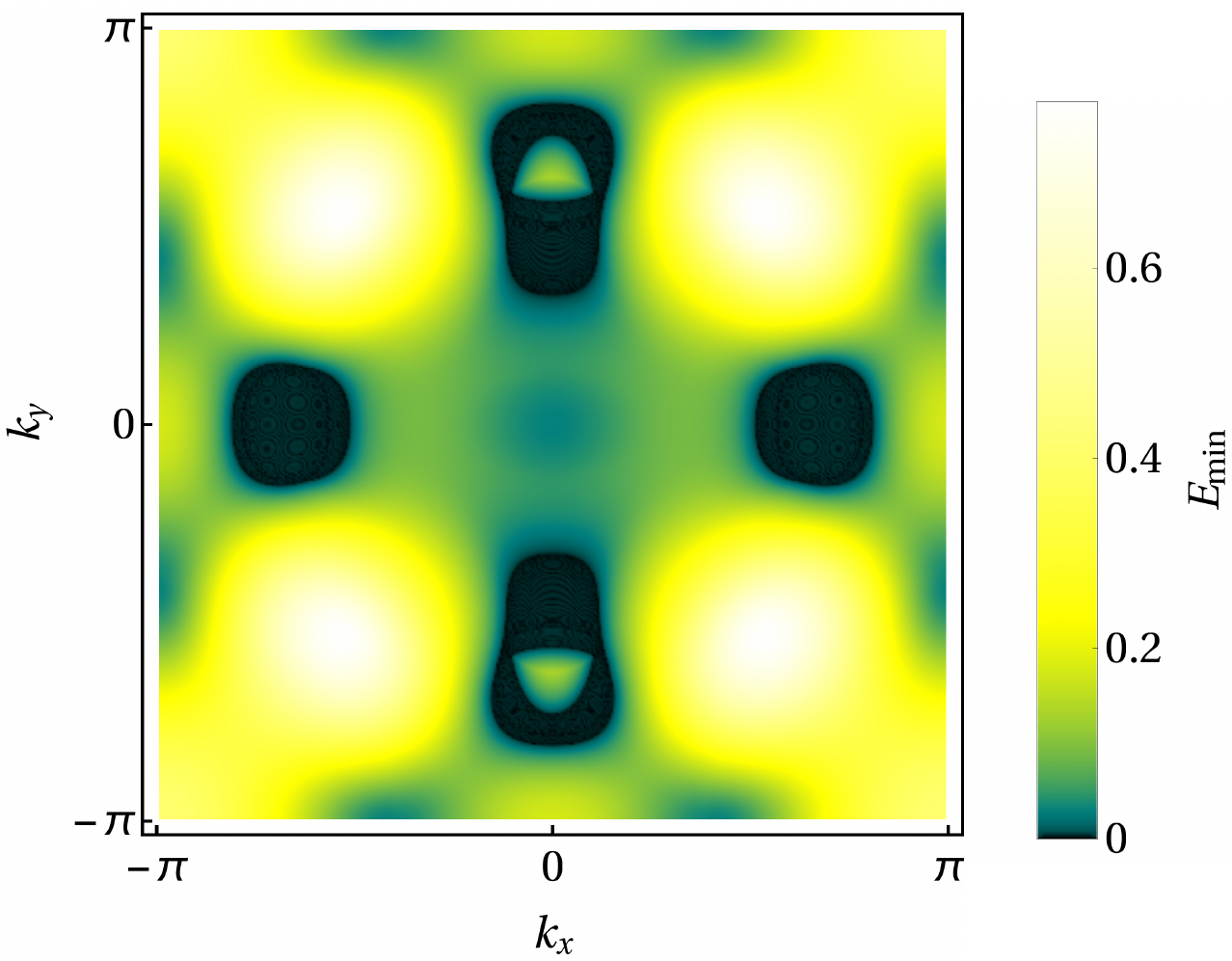}}
\caption{Smallest nonnegative eigenenergy $E_\mathrm{min}$ of the slab Hamiltonian as a function of momentum $\mathbf{k}_\|=(k_x,k_y)$ in the two-dimensional Brillouin zone. Zero-energy states are shown in black. The black regions are the projections of the BFSs. The thickness of the slab is $N=180$ unit cells.}
\label{fig.spec2D}
\end{figure}

\begin{figure}[tb]
\centerline{\includegraphics[width=0.9\columnwidth]{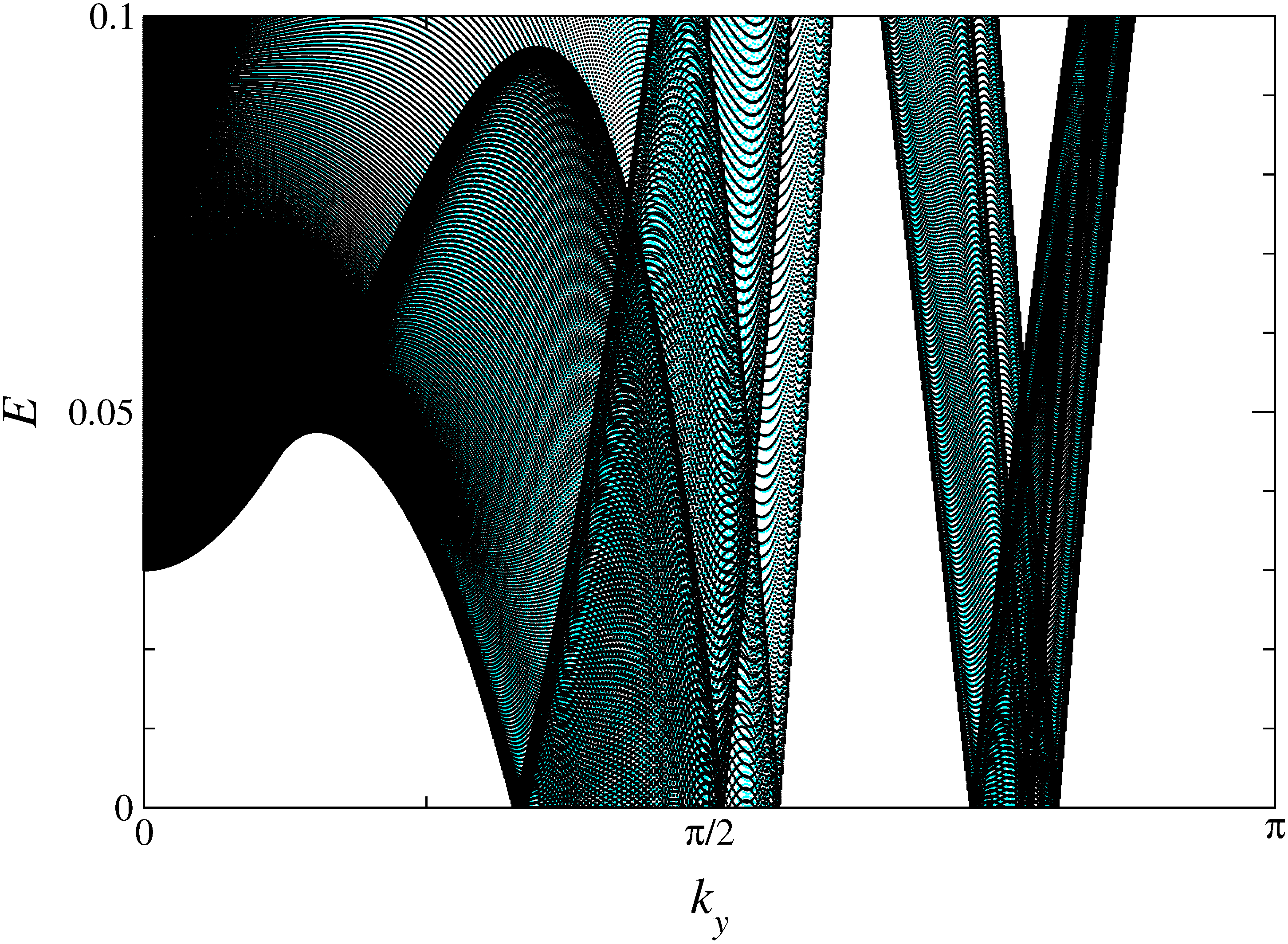}}
\caption{Dispersion of low-lying nonnegative eigenenergies of the slab Hamiltonian (black) along the $k_x=0$ axis in the two-dimensional Brillouin zone. The thickness of the slab is $N=100$. The dispersion is overlaid over the nonnegative eigenenergies for periodic boundary conditions and the same thickness (cyan), which mimic the bulk dispersion.}
\label{fig.spec2D.cut}
\end{figure}

We then Fourier transform the Bogoliubov--de Gennes Hamiltonian for the superconducting state into real space in the \textit{z} direction and introduce open boundary conditions, i.e., we describe a slab with $(001)$ surfaces. Finally, we diagonalize the slab Hamiltonian for many momenta $\mathbf{k}_\|=(k_x,k_y)$ in the two-dimensional Brillouin zone. Figure \ref{fig.spec2D} shows the smallest nonnegative quasiparticle energy for each momentum $\mathbf{k}_\|$. The projections of the BFSs are clearly seen as black regions. Note that the inside of the cylindrical pockets is visible. The plot is indistinguishable from the projection of the bulk dispersion into the $k_xk_y$ plane (not shown). This means that there is no sign of surface states. This is confirmed by plotting the full dispersion along a cut in momentum space in Fig.\ \ref{fig.spec2D.cut}. The projection of the BFSs here shows up as bands going to zero energy.

Hence we obtain a negative result: At least for a uniform pairing potential, the $\mathbb{Z}_2$ invariant of the BFSs is not associated with any surface states. On the other hand, the Bogolibov Fermi pockets that result from the inflation of point nodes are also protected by a nonzero even Chern number \cite{BrS17,BAM18}. It has been shown in Ref.\ \cite{BAM18} that these are associated with the expected number of Fermi arcs of surface states emanating from the projected pockets. These arcs are not visible in Figs.\ \ref{fig.spec2D} and \ref{fig.spec2D.cut} since inflated point nodes with opposite Chern numbers are separated by vectors $(0,0,\pm\pi)$ in Fig.\ \ref{fig.FS} and are thus projected on top of each other for the $(001)$ slab.

\section{Summary and conclusions}
\label{sec.summary}

In this paper, we have analyzed how BFSs in clean multiband superconductors affect experimental probes that are commonly used to characterize nodal superconductors. As a prerequisite, we have obtained the low-energy form of the DOS for rather general nodal structures. The DOS can be probed directly by tunneling experiments. We have then derived the low-temperature behavior of the magnetic penetration depth, the electronic contribution to the specific heat and Sommerfeld coefficient, the NMR spin-lattice relaxation rate, and the thermal conductivity. For ease of reference, we summarize the leading temperature dependences of these quantities for the most relevant types of nodes in Table \ref{tab.sum}. The table also shows expressions for a full gap for comparison.

All observables studied here should also show a characteristic dependence on magnetic field. However, such measurements are not easy to interpret since the field does not penetrate a superconductor uniformly. A nearly uniform field is realized in type-II superconductors close to the upper critical field. In this region, the magnetic-field dependence of the pairing amplitude $\Delta$ needs to be taken into account. More critically, the suitable magnetic-field range is rather small, probably precluding the extraction of a characteristic field dependence.

We have also considered the possibility of topologically protected surface states associated with the $\mathbb{Z}_2$ invariant of the BFS. However, at least under the simple assumption of a uniform pairing potential, there are no such surface states. Inflated point nodes with nonzero Chern numbers lead to Fermi arcs at the surfaces, though \cite{BAM18}. The character of these arcs is unchanged by the inflation of the nodes and thus not suitable for the discrimination between inflated and uninflated point nodes.

On the other hand, all probes listed in Table \ref{tab.sum}, with the exception of $\Delta\lambda$ in the case of linear point nodes, are, in principle, able to discriminate between all five types of nodal structures. Of course, the observability of BFSs always relies on the residual DOS being sufficiently large, which is controlled by the pseudomagnetic field $h$. Since $h$ is of second order in the interband pairing \cite{ABT17, BAM18}, it is generically small. Hence, low powers of $h$ are beneficial. We conclude that for $g=1$, the magnetic penetration depth $\Delta\lambda$ is the method of choice. Since $\Delta\lambda$ is not sensitive to inflation for $g=2$, in that case the best bets are the specific heat and Sommerfeld coefficient as well as the thermal conductivity.
We hope that this work will motivate experimentalists to look for BFSs in nodal superconductors.

\acknowledgments

We thank D. F. Agterberg, P. M. R. Brydon, A. Chernyshev, H.-H. Klauss, and M. Vojta for useful discussions. C.\,T. thanks the University of Otago for hospitality. Financial support by the Deut\-sche For\-schungs\-ge\-mein\-schaft through the Research Training Group GRK 1621, the Cluster of Excellence on Complexity and Topology in Quantum Matter ct.qmat (EXC 2147), and the Collaborative Research Center SFB 1143, project A04, is gratefully acknowledged.

\appendix

\section{Derivation of the density of states}
\label{app.dos}

In this appendix, we summarize the derivation of the quasiparticle DOS for several cases.

\subsection{Power-law point nodes}
\label{app.DOS.point}

Here, we outline the derivation of the DOS for point nodes with the power-law dispersion given in Eq.\ (\ref{eq.Eq.3}). To start with, we substitute $\tilde q_1 = (\alpha_1|\Delta_0|)^{1/m}\, q_1$, $\tilde q_2 = (\alpha_2|\Delta_0|)^{1/n}\, q_2$, and $\tilde q_3 = v_F q_3$ in Eq.\ (\ref{eq.DOS.point.2}), giving
\begin{align}
D(E) &= \frac{8}{(2\pi)^3}\,
  \frac{1}{(\alpha_1|\Delta_0|)^{1/m}\, (\alpha_2|\Delta_0|)^{1/n}\, v_F} \nonumber \\
&\quad {}\times \sum_{s_1,s_2=\pm 1} \int_0^\infty d\tilde q_1\, d\tilde q_2\, d\tilde q_3
  \nonumber \\
&\quad {}\times \delta\left( E - \left[s_1 h + s_2\, \sqrt{\tilde q_1^{2m} + \tilde q_2^{2n}
    + \tilde q_3^2}\right] \right) .
\end{align}
Next, the substitution $k_1 = \tilde q_1^m$, $k_2 = \tilde q_2^n$, $k_3 = \tilde q_3$ brings the dispersion into a more familiar symmetric form,
\begin{align}
D(E) &= \frac{8}{(2\pi)^3}\,
  \frac{1}{(\alpha_1|\Delta_0|)^{1/m}\, (\alpha_2|\Delta_0|)^{1/n}\, v_F} \nonumber \\
&\quad {}\times \sum_{s_1,s_2=\pm 1} \int_0^\infty dk_1\, dk_2\, dk_3\,
  \frac{1}{mn}\, k_1^{1/m-1}\, k_2^{1/n-1} \nonumber \\
&\quad {}\times \delta\left( E - \left[s_1 h + s_2\, \sqrt{ k_1^2 + k_2^2 + k_3^2 } \right] \right) .
\end{align}
The symmetric dispersion suggests to introduce spherical coordinates for $(k_1,k_2,k_3)$, giving
\begin{align}
D(E) &= \frac{8}{(2\pi)^3}\,
  \frac{1}{mn\, (\alpha_1|\Delta_0|)^{1/m}\, (\alpha_2|\Delta_0|)^{1/n}\, v_F} \nonumber \\
&\quad {}\times \sum_{s_1,s_2=\pm 1} \int_0^\infty dk \int_0^{\pi/2} d\theta
  \int_0^{\pi/2} d\phi\, k^{1/m+1/n} \nonumber \\
&\quad {}\times \sin^{1/m+1/n-1}\!\theta\, \cos^{1/m-1}\!\phi\, \sin^{1/n-1}\!\phi \nonumber \\
&\quad {}\times \delta( E - [s_1 h + s_2 k] ) \nonumber \\
&= \frac{8}{(2\pi)^3}\,
  \frac{1}{mn\, (\alpha_1|\Delta_0|)^{1/m}\, (\alpha_2|\Delta_0|)^{1/n}\, v_F} \nonumber \\
&\quad {}\times \big( |E + h|^{1/m+1/n} + |E - h|^{1/m+1/n} \big) \nonumber \\
&\quad {}\times \frac{\sqrt{\pi}\: \Gamma\left( \frac{1}{2m} + \frac{1}{2n} \right)}
    {2 \Gamma\left( \frac{1}{2} + \frac{1}{2m} + \frac{1}{2n} \right)}\,
    \frac{\Gamma\left(\frac{1}{2m}\right) \Gamma\left(\frac{1}{2n}\right)}
      {2\, \Gamma\left(\frac{1}{2m} + \frac{1}{2n}\right)} \nonumber \\
&= \frac{2\sqrt{\pi}}{(2\pi)^3}\,
  \frac{1}{mn\, (\alpha_1|\Delta_0|)^{1/m}\, (\alpha_2|\Delta_0|)^{1/n}\, v_F} \nonumber \\
&\quad {}\times \big( |E + h|^{1/m+1/n} + |E - h|^{1/m+1/n} \big) \nonumber \\
&\quad {}\times \frac{\Gamma\left(\frac{1}{2m}\right) \Gamma\left(\frac{1}{2n}\right)}
    {\Gamma\left(\frac{1}{2} + \frac{1}{2m} + \frac{1}{2n}\right)} .
\end{align}

\subsection{Generalized double-Weyl point nodes}
\label{app.DOS.Weyl}

In the following, we show the derivation of the DOS for the general quadratic dispersion in Eq.\ (\ref{eq.Eq.5}). The substitution $\tilde q_1 = \sqrt{\alpha_1|\Delta_0|}\, q_1$, $\tilde q_2 = \sqrt{\alpha_2|\Delta_0|}\, q_2$, $\tilde q_3 = v_F q_3$ in the DOS, Eq.\ (\ref{eq.DOS.point.2}), gives
\begin{align}
D(E) &= \frac{8}{(2\pi)^3}\,
  \frac{1}{\sqrt{\alpha_1 \alpha_2}\, |\Delta_0|\, v_F}
  \sum_{s_1,s_2=\pm 1} \int_0^\infty d\tilde q_1\, d\tilde q_2\, d\tilde q_3 \nonumber \\
&{}\times \delta\left( E - \left[ s_1 h + s_2\, \sqrt{\tilde q_1^4 + \tilde q_2^4
    + 2\alpha\, \tilde q_1^2 \tilde q_2^2 + \tilde q_3^2} \right] \right) ,
\end{align}
where $\alpha \equiv \alpha_{12}^2/\alpha_1 \alpha_2$. The case $\alpha_1 = \alpha_2 = \alpha_{12}$, i.e., $\alpha=1$, corresponds to isotropic double-Weyl nodes, whereas $\alpha_{12}=0$ ($\alpha=0$) is the special case $m=n=2$ of our previous derivation. The substitution $k_1 = \tilde q_1^2$, $k_2 = \tilde q_2^2$, $k_3 = \tilde q_3$ brings the dispersion into more familiar form,
\begin{align}
D(E) &= \frac{2}{(2\pi)^3}\,
  \frac{1}{\sqrt{\alpha_1 \alpha_2}\, |\Delta_0|\, v_F} \nonumber \\
&{}\times \sum_{s_1,s_2=\pm 1} \int_0^\infty dk_1\, dk_2\, dk_3  \frac{1}{\sqrt{k_1 k_2}}
  \nonumber \\
&{}\times \delta\left( E - \left[ s_1 h + s_2\, \sqrt{k_1^2 + k_2^2 + 2\alpha k_1k_2
    + k_3^2} \right] \right) .
\end{align}
Another substitution
\begin{align}
\tilde k_1 &= \sqrt{\frac{1 + \sqrt{1-\alpha^2}}{2}}\: k_1
  + \sqrt{\frac{1 - \sqrt{1-\alpha^2}}{2}}\: k_2 , \\
\tilde k_2 &= \sqrt{\frac{1 - \sqrt{1-\alpha^2}}{2}}\: k_1
  + \sqrt{\frac{1 + \sqrt{1-\alpha^2}}{2}}\: k_2 , \\
\tilde k_3 &= k_3
\end{align}
gives
\begin{align}
D(E) &= \frac{2}{(2\pi)^3}\,
  \frac{1}{\sqrt{\alpha_1 \alpha_2}\, |\Delta_0|\, v_F} \nonumber \\
&\quad {}\times \sum_{s_1,s_2=\pm 1} \int_0^\infty d\tilde k_3 \int d^2\tilde k\,
  \frac{1}{\sqrt{ \tilde k_1 \tilde k_2
    - \frac{\alpha}{2}\, (\tilde k_1^2 + \tilde k_2^2) }} \nonumber \\
&\quad {}\times \delta\left( E - \left[ s_1 h + s_2\,
    \sqrt{\tilde k_1^2 + \tilde k_2^2 + \tilde k_3^2} \right] \right) ,
\end{align}
where the limits of the in-plane integral $\int d^2\tilde k$ are such that
\begin{equation}
k_1 k_2 = \frac{\tilde k_1\tilde k_2
  - \frac{\alpha}{2}\, (\tilde k_1^2 + \tilde k_2^2)}
  {1-\alpha^2} \ge 0 .
\end{equation}
These limits are simplified by introducing polar coordinates $\rho$, $\phi$ for $(\tilde k_1, \tilde k_2)$, giving
\begin{align}
D(E) &= \frac{2\sqrt{2}}{(2\pi)^3}\,
  \frac{1}{\sqrt{\alpha_1 \alpha_2}\, |\Delta_0|\, v_F} \nonumber \\
&\quad {}\times \sum_{s_1,s_2=\pm 1} \int_0^\infty d\tilde k_3
  \int_0^\infty d\rho \int_{\phi_-}^{\phi_+} d\phi\,
  \frac{1}{\sqrt{ \sin 2\phi - \alpha }} \nonumber \\
&\quad {}\times \delta\left( E - \left[ s_1 h + s_2\, \sqrt{\rho^2 + \tilde k_3^2} \right] \right) ,
\end{align}
with $\phi_- = (\arcsin \alpha)/2$ and $\phi_+ = (\pi - \arcsin \alpha)/2$. Note that the limits of the angular integral do not depend on the other integration variables. The angular integration decouples and can be evaluated as 
\begin{align}
G(\alpha) &\equiv \int_{\phi_-}^{\phi_+} \frac{d\phi}{\sqrt{ \sin 2\phi - \alpha }}
  = -\frac{1}{\sqrt{1-\alpha}} \nonumber \\
&{}\times \left[ F\left( \frac{\pi}{4} - \phi_+, \frac{2}{1-\alpha} \right)
    - F\left( \frac{\pi}{4} - \phi_-, \frac{2}{1-\alpha} \right) \right] ,
\end{align}
where $F$ is the incomplete elliptic integral of the first kind. The function  $G(\alpha)$ is positive and bounded for $\alpha\in[0,1)$. After introducing polar coordinates once more, this time in the $(\rho,\tilde k_3)$ plane, the $\delta$ distribution can easily be integrated over, giving
\begin{equation}
D(E) = \frac{\pi\sqrt{2}}{(2\pi)^3}\,
  \frac{1}{\sqrt{\alpha_1 \alpha_2}\, |\Delta_0|\, v_F}\, G(\alpha)\,
  \big( |E + h| + |E - h| \big) .
\end{equation}

\subsection{Power-law line nodes}
\label{app.DOS.line}

In this appendix, we derive the DOS for a circular line node with power-law dispersion, given in Eq.\ (\ref{eq.Eq.6}). Performing the integration along the line node and substituting $\tilde q_2 = (\alpha_2 |\Delta_0|)^{1/n}\, q_2$, $\tilde q_3 = v_F q_3$ in the DOS, Eq.\ (\ref{eq.DOS.point.2}), gives
\begin{align}
D(E) &= \frac{8\pi}{(2\pi)^3}\, \frac{k_F}{(\alpha_2 |\Delta_0|)^{1/n}\, v_F}
  \sum_{s_1,s_2=\pm 1} \int_0^\infty d\tilde q_2\, d\tilde q_3 \nonumber \\
&\quad {}\times
  \delta\left( E - \left[ s_1 h + s_2\, \sqrt{\tilde q_2^{2n} + \tilde q_3^2} \right] \right) .
\end{align}
Another substitution, $k_2 = \tilde q_2^n$, $k_3 = \tilde q_3$, brings the dispersion into more familiar form,
\begin{align}
D(E) &= \frac{8\pi}{(2\pi)^3}\, \frac{k_F}{(\alpha_2 |\Delta_0|)^{1/n}\, v_F}
  \sum_{s_1,s_2=\pm 1} \int_0^\infty dk_2\, dk_3 \nonumber \\
&\quad {}\times \frac{1}{n}\, k_2^{1/n-1}\,
  \delta\left( E - \left[ s_1 h + s_2\, \sqrt{k_2^2 + k_3^2} \right] \right) ,
\end{align}
and with polar coordinates $\rho$, $\phi$ for $(k_2,k_3)$ we obtain
\begin{align}
D(E) &= \frac{8\pi}{(2\pi)^3}\, \frac{k_F}{n\, (\alpha_2 |\Delta_0|)^{1/n}\, v_F} \nonumber \\
&\quad {}\times \sum_{s_1,s_2=\pm 1} \int_0^\infty d\rho \int_0^{\pi/2} d\phi\, \rho^{1/n}
  \cos^{1/n-1}\!\phi \nonumber \\
&\quad {}\times \delta( E - [s_1 h + s_2 \rho] ) \nonumber \\
&= \frac{4\pi^{3/2}}{(2\pi)^3}\, \frac{k_F}{n\, (\alpha_2 |\Delta_0|)^{1/n}\, v_F} \nonumber \\
&\quad {}\times \big( |E + h|^{1/n} + |E - h|^{1/n} \big)\,
  \frac{\Gamma\left(\frac{1}{2n}\right)}
    {\Gamma\left(\frac{1}{2} + \frac{1}{2n}\right)} .
\end{align}

\section{Derivation of the thermal conductivity}
\label{app.thermal}

In this appendix, we derive the thermal conductivity for linearly dispersing point and line nodes.

\subsection{Linear point nodes}
\label{app.thermal.point}

We first consider point nodes. The linear quasiparticle dispersion in Eq.\ (\ref{eq.Eq.4}) yields two bands at $E>0$. For both of them we find
\begin{equation}
\frac{\partial E}{\partial \mathbf{q}} \otimes \frac{\partial E}{\partial \mathbf{q}}
  = \frac{\mathcal{V}}{\alpha_1^2 |\Delta_0|^2\, q_1^2
    + \alpha_2^2 |\Delta_0|^2\, q_2^2 + v_F^2\, q_3^2} ,
\end{equation}
where the symmetric tensor $\mathcal{V}$ has the components
\begin{align}
\mathcal{V}_{11} &= \alpha_1^4 |\Delta_0|^4\, q_1^2 , \\
\mathcal{V}_{12} &= \alpha_1^2 \alpha_2^2 |\Delta_0|^4\, q_1 q_2 , \\
\mathcal{V}_{13} &= \alpha_1^2 |\Delta_0|^2 v_F^2\, q_1 q_3 , \\
\mathcal{V}_{22} &= \alpha_2^4 |\Delta_0|^4\, q_2^2 , \\
\mathcal{V}_{23} &= \alpha_2^2 |\Delta_0|^2 v_F^2\, q_2 q_3 , \\
\mathcal{V}_{33} &= v_F^4\, q_3^2 .
\end{align}
With the substitution $\tilde q_1 = \alpha_1 |\Delta_0|\, q_1$, $\tilde q_2 = \alpha_2 |\Delta_0|\, q_2$, $\tilde q_3 = v_F q_3$, we can write the average for one of the bands as
\begin{equation}
\left\langle \frac{\partial E}{\partial \mathbf{q}} \otimes \frac{\partial E}{\partial \mathbf{q}}
  \right\rangle_{\!E}
  = \frac{\displaystyle \int d^3\tilde q\, \frac{\mathcal{V}}{\tilde q^2}\,
      \delta( E \mp h - |\tilde{\mathbf{q}}| )}
    {\displaystyle \int d^3\tilde q\, \delta( E \mp h - |\tilde{\mathbf{q}}| )} .
\end{equation}
The off-diagonal components of $\mathcal{V}$ vanish upon integration since they are odd in momentum components. The diagonal ones can easily be evaluated in spherical coordinates, giving
\begin{equation}
\left\langle \frac{\partial E}{\partial \mathbf{q}} \otimes \frac{\partial E}{\partial \mathbf{q}}
  \right\rangle_{\!E}
  = \left( \begin{array}{ccc}
    \displaystyle \frac{\alpha_1^2 |\Delta_0|^2}{3} & 0 & 0 \\
    0 & \displaystyle \frac{\alpha_2^2 |\Delta_0|^2}{3} & 0 \\
    0 & 0 & \displaystyle \frac{v_F^2}{3}
  \end{array} \right) .
\end{equation}
This average is the same for both bands and is in fact independent of the pseudomagnetic field and of the energy. Hence, the thermal conductivity is
\begin{equation}
\kappa = c\tau \left( \begin{array}{ccc}
    \displaystyle \frac{\alpha_1^2 |\Delta_0|^2}{3} & 0 & 0 \\
    0 & \displaystyle \frac{\alpha_2^2 |\Delta_0|^2}{3} & 0 \\
    0 & 0 & \displaystyle \frac{v_F^2}{3}
  \end{array} \right) ,
\end{equation}
where $c$ is the specific heat, which is given in Eq.\ (\ref{eq.c.8}).

\subsection{Linear line nodes}
\label{app.thermal.line}

For a linearly dispersing circular line node in the $k_xk_y$ plane, the tensor product of velocities is
\begin{equation}
\frac{\partial E}{\partial \mathbf{k}} \otimes \frac{\partial E}{\partial \mathbf{k}}
  = \frac{\mathcal{V}}{ v_F^2\, (k_\rho-k_F)^2 + \alpha_2^2 |\Delta_0|^2\, k_z^2 } ,
\end{equation}
where $\mathcal{V}$ now has the Cartesian components
\begin{align}
\mathcal{V}_{xx} &= v_F^4\, (k_\rho-k_F)^2\, \cos^2\! \phi , \\
\mathcal{V}_{xy} &= v_F^4\, (k_\rho-k_F)^2\, \cos\phi \sin\phi , \\
\mathcal{V}_{xz} &= \alpha_2^2 |\Delta_0|^2 v_F^2\, (k_\rho-k_F)\, k_z \cos\phi , \\
\mathcal{V}_{yy} &= v_F^4\, (k_\rho-k_F)^2\, \sin^2\! \phi , \\
\mathcal{V}_{yz} &= \alpha_2^2 |\Delta_0|^2 v_F^2\, (k_\rho-k_F)\, k_z \sin\phi , \\
\mathcal{V}_{zz} &= \alpha_2^4 |\Delta_0|^4\, k_z^2 .
\end{align}
Here, $k_\rho$, $\phi$, and $k_z$ are the cylindrical coordinates describing the momentum $\mathbf{k}$. The average over $\phi$ gives $\langle \cos^2\!\phi \rangle_E = \langle \sin^2\!\phi \rangle_E = 1/2$, whereas all other functions of $\phi$ average to zero. For the remaining integrals, we substitute $\tilde q_2 = v_F\, (k_\rho-k_F)$ and $\tilde q_3 = \alpha_2 |\Delta_0|\, k_z$. The average then reads
\begin{equation}
\left\langle \frac{\partial E}{\partial \mathbf{k}} \otimes \frac{\partial E}{\partial \mathbf{k}}
  \right\rangle_{\!E}
  = \frac{\displaystyle \int d^2\tilde q\, \frac{\tilde{\mathcal{V}}}{\tilde q^2}\,
      \delta(E \mp h - |\tilde{\mathbf{q}}|)}
    {\displaystyle \int d^2\tilde q\, \delta(E \mp h - |\tilde{\mathbf{q}}|)} ,
\end{equation}
with
\begin{align}
\tilde{\mathcal{V}}_{xx} &=
  \tilde{\mathcal{V}}_{yy} = \frac{1}{2}\, v_F^2\, \tilde q_2^2 , \\
\tilde{\mathcal{V}}_{zz} &= \alpha_2^2 |\Delta_0|^2\, \tilde q_3^2 ,
\end{align}
and the off-diagonal components are zero. Once again introducing polar coordinates, this time for $(\tilde q_2,\tilde q_3)$, we find
\begin{equation}
\left\langle \frac{\partial E}{\partial \mathbf{q}} \otimes \frac{\partial E}{\partial \mathbf{q}}
  \right\rangle_{\!E}
  = \left( \begin{array}{ccc}
    \displaystyle \frac{v_F^2}{4} & 0 & 0 \\
    0 & \displaystyle \frac{v_F^2}{4} & 0 \\
    0 & 0 & \displaystyle \frac{\alpha_2^2 |\Delta_0|^2}{2}
  \end{array} \right) ,
\end{equation}
which is again energy independent, and thus
\begin{equation}
\kappa = c\tau \left( \begin{array}{ccc}
    \displaystyle \frac{v_F^2}{4} & 0 & 0 \\
    0 & \displaystyle \frac{v_F^2}{4} & 0 \\
    0 & 0 & \displaystyle \frac{\alpha_2^2 |\Delta_0|^2}{2}
  \end{array} \right) .
\end{equation}
The specific heat $c$ is given by Eq.~(\ref{eq.c.10}).


\begin{thebibliography}{99}
\bibitem{note.full}The results for a constant gap can be obtained by inserting the BCS form of the DOS, $D(E)=0$ for $E<\Delta$ and $D(E) \propto E/\sqrt{E^2-\Delta^2}$ for $E>\Delta$, into Eqs.\ (\ref{eq.lambda.2}), (\ref{eq.c.2}), and (\ref{eq.T1.2}) and in the last case introducing the nuclear resonance frequency by replacing $D^2(E)$ with $D(E)\,D(E+\omega_0)$.
\bibitem{note.negative}Note that the terms in Eq.\ (\ref{eq.lambda.4}) are also correct for $g-2j<0$.
\bibitem{note.irrep}The time-reversal-symmetry-breaking pairing that is required to obtain BFSs is most naturally realized in a mul\-ti-di\-men\-sio\-nal irreducible representation.
\end{thebibliography}
\end{document}